\documentclass[letter]{aa}

\usepackage[flushleft]{threeparttable}
\usepackage{graphicx}
\usepackage{xtab,afterpage}
\usepackage{amsmath}
\usepackage{epstopdf}
\usepackage{url}
\usepackage[flushleft]{threeparttable}
\usepackage{booktabs,fixltx2e}
\usepackage{amssymb}
\usepackage{textcomp}
\usepackage{hyperref} 
\usepackage{multirow} 
\usepackage{times}
\usepackage{txfonts}
\usepackage{footnote}
\usepackage{booktabs,caption,fixltx2e}
\usepackage[flushleft]{threeparttable}

\makeatletter
\newcommand*{\rom}[1]{\expandafter\@slowromancap\romannumeral #1@}
\makeatother

\title{A molecular wind blows out of the Kuiper belt}
\author{Quentin Kral\inst{1}\thanks{E-mail: quentin.kral@obspm.fr} \and J. E. Pringle\inst{2} \and Aurélie Guilbert-Lepoutre\inst{3} \and Luca Matr{\`a}\inst{4} \and Julianne I. Moses\inst{5} \and Emmanuel Lellouch\inst{1} \and Mark C. Wyatt\inst{2} \and Nicolas Biver\inst{1} \and Dominique Bockelée-Morvan\inst{1} \and Amy Bonsor\inst{2}, Franck Le Petit\inst{6} \and G. Randall Gladstone\inst{7,8}}

\institute{LESIA, Observatoire de Paris, Universit{\'e} PSL, CNRS, Sorbonne Universit{\'e}, Univ. Paris Diderot, Sorbonne Paris Cit{\'e}, 5 place Jules Janssen, 92195 Meudon, France\\
\and
Institute of Astronomy, University of Cambridge, Madingley Road, Cambridge CB3 0HA, UK\\
\and
LGL-TPE, UMR 5276, CNRS, Claude Bernard Lyon 1 University, ENS Lyon, Villeurbanne Cedex, France\\
\and
School of Physics, National University of Ireland Galway, University Road, Galway, Ireland\\
\and
Space Science Institute, 4765 Walnut St, Suite B, Boulder, CO 80301, USA\\
\and
LERMA, Observatoire de Paris, Université PSL, CNRS, Sorbonne Université, Univ. Paris Diderot,\\ Sorbonne Paris Cité, 5 place Jules Janssen, 92195 Meudon, France\\
\and
Southwest Research Institute, San Antonio, TX 78238, United States
\and
The University of Texas at San Antonio, San Antonio TX 78249, United States
}

\begin{document}

   \date{Received September 15, 1932; accepted March 16, 1937}



\label{firstpage}


  \abstract
   {Gas has been detected in many exoplanetary systems ($>$10 Myr), thought to be released in the destruction of volatile-rich planetesimals orbiting in exo-Kuiper belts.}
   {In this letter, we aim to explore whether gas is also expected in the Kuiper belt (KB) in our Solar System.}
   {To quantify the gas release in our Solar System, we use models for gas release that have been applied to extrasolar planetary systems, as well as a physical model that accounts for gas released due to the progressive internal warming of large planetesimals.}
   {We find that only bodies larger than about 4 km can still contain CO ice after 4.6 Gyr of evolution. This finding may provide a clue as to why Jupiter-family comets, thought to originate in the Kuiper belt, are deficient in CO compared to Oort-clouds comets. We predict that gas is still produced in the KB right now at a rate of $2 \times 10^{-8}$ M$_\oplus$/Myr for CO and orders of magnitude more when the Sun was younger. Once released, the gas is quickly pushed out by the Solar wind. Therefore, we predict a gas wind in our Solar System starting at the KB location and extending far beyond with regards to the heliosphere with a current total CO mass of $\sim 2 \times 10^{-12}$ M$_\oplus$ (i.e. 20 times the CO quantity that was lost by the Hale-Bopp comet during its 1997 passage) and CO density in the belt of $3 \times 10^{-7}$ cm$^{-3}$. We also predict the existence of a slightly more massive atomic gas wind made of carbon and oxygen (neutral and ionized) with a mass of $\sim 10^{-11}$ M$_\oplus$.}
   {We predict that gas is currently present in our Solar System beyond the Kuiper belt and that although it cannot be detected with current instrumentation, it could be observed in the future with an in situ mission using an instrument similar to Alice on New Horizons with larger detectors. Our model of gas release due to slow heating may also work for exoplanetary systems and provide the first real physical mechanism for the gas observations. Lastly, our model shows that the amount of gas in the young Solar System should have been orders of magnitude greater and that it may have played an important role for, e.g., planetary atmosphere formation.}


   \maketitle
   
\section{Introduction}
The past decade was very prolific in terms of detecting gas (mostly CO, C and O) around main sequence stars, therefore changing the paradigm of evolved planetary systems that were thought to be devoid of gas after 10 Myr. Indeed, most bright exoplanetesimal belts show the presence of gas, as demonstrated recently with ALMA \citep{2017ApJ...849..123M}, and it could be that all these belts have gas at some level (even if undetectable with current instruments). These belts, similar to our Kuiper belt, are made of large bodies colliding with each other and creating dust that can then be observed around extrasolar stars through its emission in the infrared above that of the star, which can be resolved at high resolution (showing, e.g., gaps and asymmetries that may be related to the presence of planets). 

Recent models show that the best explanation for the CO gas observed co-located with exo-Kuiper belts is a secondary production (i.e., the gas is not a remnant of the young planet-forming disks that persist for less than 10 Myr), where CO is released from planetesimals at a rate proportional to their collisional frequency \citep{2016MNRAS.461..845K,2017MNRAS.469..521K}. Observations of carbon and oxygen atoms are explained as the daughter species of CO photodissociation within the framework of this model. The most massive gas disks with CO masses close to the amount observed in younger planet-forming disks have first been called hybrid disks \citep[e.g.,][because the gas was thought to be primordial but the dust secondary]{2013ApJ...776...77K}. We can now also explain the gas in these previously considered hybrid disks as entirely secondary, because CO released from planetesimals becomes shielded by the carbon produced when it photodissociates (and by CO itself through self-shielding), which can then accumulate to large amounts \citep{2019MNRAS.489.3670K,2020MNRAS.492.4409M}.

In addition, recent observations showed that comets in our Solar System start being active as far as the Kuiper belt distance. The long period comet (3 Myr) C/2017 K2 (PANSTARRS) exhibited activity as far as 9-16 au, and models show that dust production (presumably driven by sublimation of CO) needs to have started at the KB (35 au) to explain the photometric data \citep{2021arXiv210206313J}. Historically, there are also other comets showing distant CO outgassing such as C/1995 O1 \citep[Hale-Bopp,][]{2002EM&P...90....5B}, or the short-period comets (Centaur-like) 29P/Schwassmann-Wachmann 1 and 2060 Chiron \citep{2017PASP..129c1001W}.

Given all this new knowledge in terms of CO outgassing in comets and exocomets, we want to explore what it means for our own Solar System. For instance, what gas production rate do we predict for the current Kuiper belt? Would sublimation still be active enough in releasing CO as far as the KB distance that it can be detected? Is the dynamics of the released gas around our G2-type Sun similar to that previously observed (predominantly though not exclusively) around young main-sequence stars? If gas is released in the KB, how can we detect it and how does it affect the system as a whole? These are the questions we are tackling in this paper.

\section{Results}

\begin{figure}
   \centering
   \includegraphics[width=9.5cm]{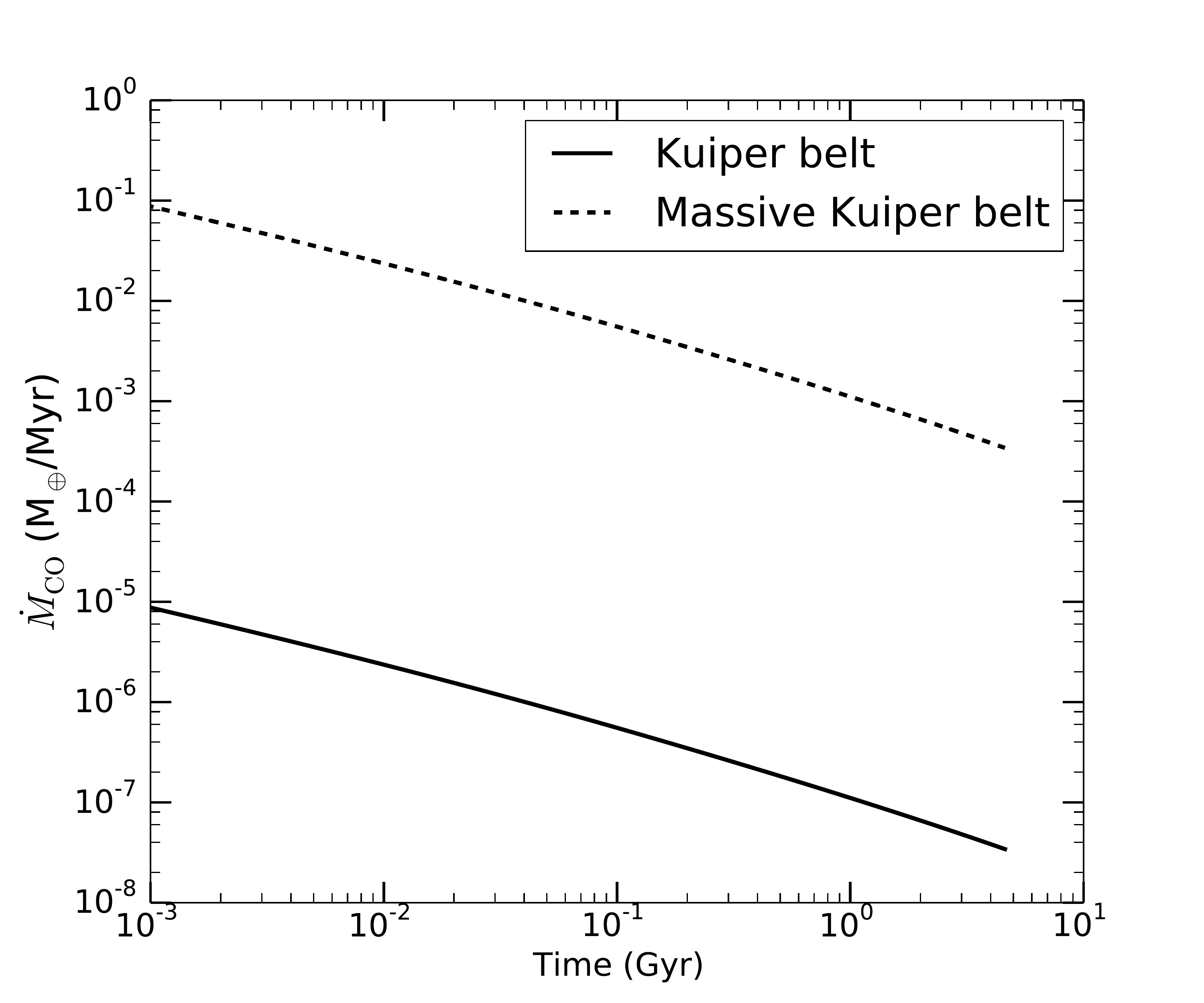}
   \caption{\label{fig1} CO gas production rate in the Kuiper belt $\dot{M}_{\rm CO}$ as a function of time ($t=0$ is when the gas release starts, i.e., probably after a few Myr to 10 Myr and the end of the lines on the right is today) predicted by our sublimation model (see Appendix \ref{subl}). The solid line is for the Kuiper belt assuming it starts with a low mass similar to the current KB mass and the dashed line is for a more massive belt similar to the archetype $\beta$ Pic belt.}
\end{figure}

To answer these questions, we first extrapolate the gas production rate in our KB from the most recent extrasolar models. To do so, we compute the dust mass loss rate in the KB due to collisions using a state-of-the-art model of dust in our Solar System \citep{2012A&A...540A..30V,2021Icar..35614256M}. Indeed, according to extrasolar models that fit most observations to date \citep{2017MNRAS.469..521K}, the gas production rate is proportional to the mass loss rate of the belt's collisional cascade, and we find (see Appendix \ref{gasprod}) that $\sim 10^{-9}$ M$_\oplus$/Myr of CO gas should be released in the current KB. The model's idea (which fits extrasolar observations) is that large planetesimals are composed of $\sim$10\% of CO \citep[see][]{2017MNRAS.469..521K} that is released along with collisions that produce the observed dust (but the detailed physical mechanism is not constrained), either at the top (large km bodies) or further down the collisional cascade, but before solid bodies are ground down to dust and expelled by radiation pressure. We also use a more direct approach relying on the counting rate of the New Horizons dust counter to determine the dust production rate (rather than a numerical model) and arrive at the same value for the gas production rate. We also test a different more physically motivated model for releasing the CO and assume it comes from the slow heating provided by the Sun over long timescales, which warms up large bodies at greater depths as time goes by, and releases the CO in these increasingly deeper layers. We find that after 4.6 Gyr of evolution, only bodies larger than about 4 km can still contain CO (smaller bodies would have lost it already), and all together they release CO at a rate of $\sim 2 \times 10^{-8}$ M$_\oplus$/Myr. In this model, a single 30 km radius planetesimal would release around $10^{-14}$ M$_\oplus$/Myr, i.e., it is much lower than what can be detected with missions targeting specific KBOs \citep[e.g.,][]{2021Icar..35614072L}. Fig.~\ref{fig1} shows the temporal evolution of the release rate, which goes down with time as only larger and larger bodies can participate as time goes by (see Appendix \ref{subl}). We note that this means that sampling the material in the KB now would not lead to the primordial volatile composition of planetesimals. We also test this slow stellar-driven heating model on more massive belts (similar to those observed around extrasolar systems) and show that it provides the right order of magnitude to explain CO observed around younger exo-systems, which may provide the first physical explanation for their ubiquitous CO presence.

\begin{figure}
   \centering
   \includegraphics[width=8.cm]{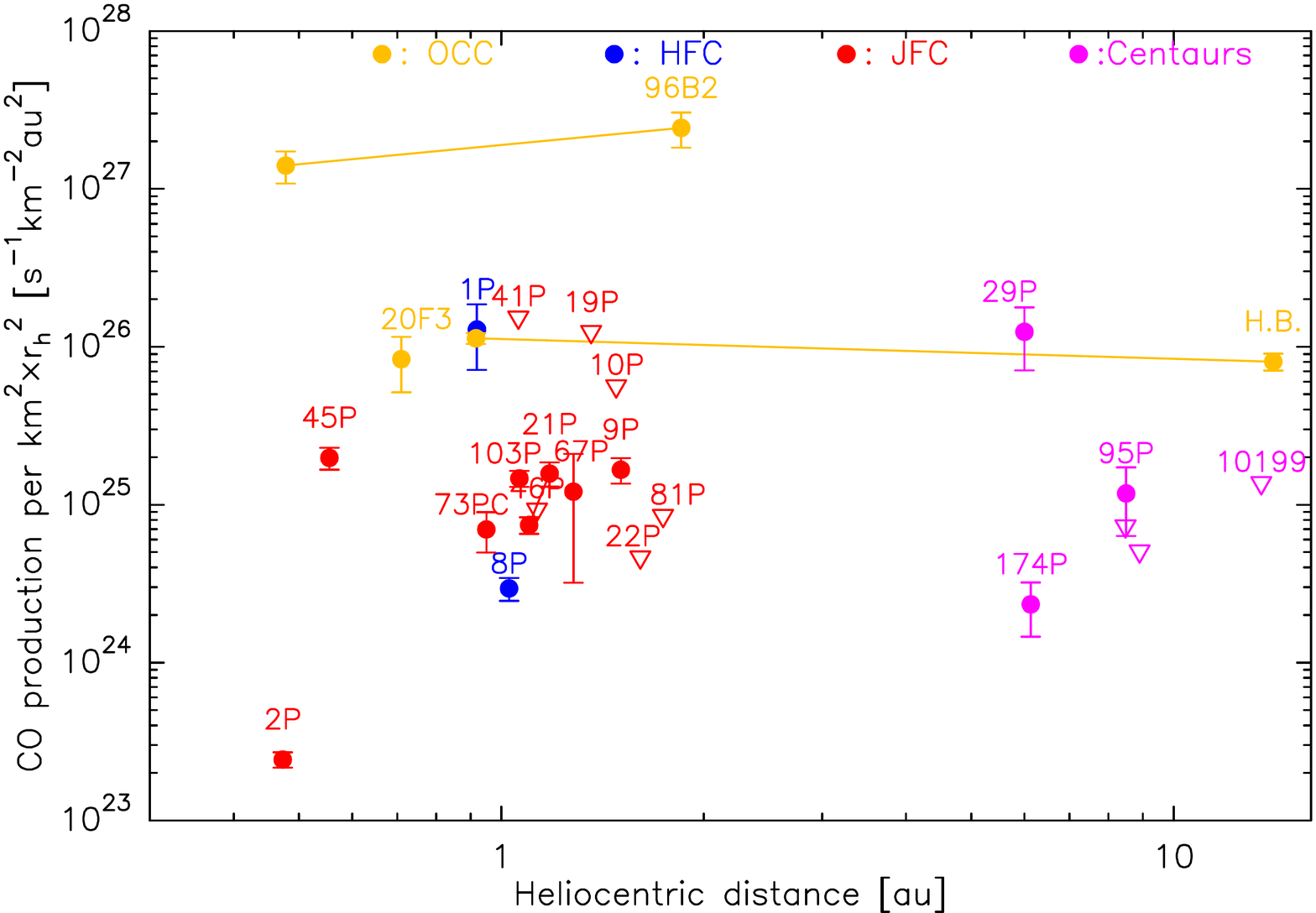}
      \includegraphics[width=8.cm]{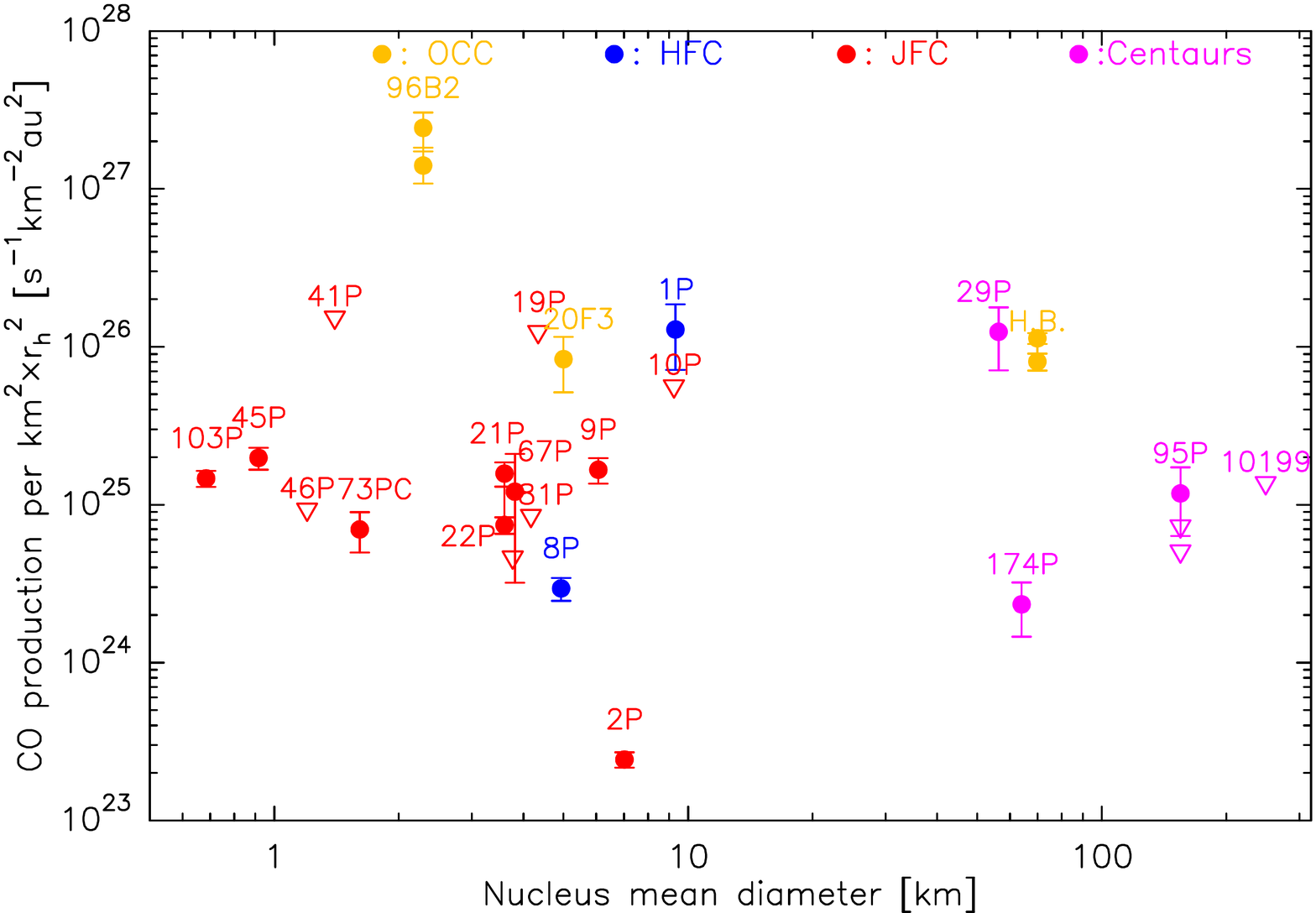}
   \caption{\label{figjfc} CO production rate per km$^2$ for Oort cloud comets (OCCs, orange), Halley family comets (HFCs, blue), Jupiter family comets (JFCs, red) and Centaurs (purple) of diameter $D$ scaled by the heliocentric distance $r_h$ squared - or in other terms $(Q_{\rm CO}/D^2) r_h^2$ - as a function of $r_h$ (top) and $D$ (bottom). Downwards triangles are for upper limits. H.B. stands for Hale-Bopp, 20F3 for C/2020 F3 Neowise, 96B2 for C/1996 B2 Hyakutake and other objects have their full names. The data are listed in Table~\ref{tabcom}.}
\end{figure}

Comets show a diversity in composition, with a factor of 10-100 variability in the volatile abundance \citep{2017RSPTA.37560252B}. However, in general this diversity does not appear correlated with the dynamical category. An exception is for CO, whose abundance relative to water appears depleted in Jupiter family comets (JFC), by a factor $\sim$4 in average compared to Oort Cloud comets \citep{2016Icar..278..301D}. As we show in Fig.~\ref{figjfc}, the CO depletion in JFC compared to other comets is also visible if expressed as a specific CO production rate, i.e., the production rate per unit area $Q_{\rm CO}/(\pi D^2/4)$, where $D$ is the equivalent diameter. The top panel of Fig.~\ref{figjfc} shows this specific production rate, multiplied by $r_h^2$, as a function of the heliocentric distance $r_h$ of the measurements. As demonstrated for C/1996 B2 Hyakutake
and C/1995 O1 Hale-Bopp (96B2 and H.B. in the figure), the scaling by $r_h^2$ corrects to first order for the distance effects and allows comparison of measurements at different distances. In the bottom panel of Fig.~\ref{figjfc}, this distance-corrected specific CO production rate is shown as a function of $D$. In both panels of Fig.~\ref{figjfc}, JFCs clearly appear CO-depleted with respect to Oort-cloud comets (OCCs).
As JFCs are thought to originate from the transneptunian region, in particular the Scattered Disk \citep[e.g.,][]{2004come.book..193D,2020SSRv..216....6W} and most of them have diameters $<$ 5 km, our calculation that only Kuiper belt bodies larger than 4 km can retain CO over the
age of the Solar System may provide a natural explanation to this behaviour. Interestingly, the observed cumulative size distribution of JFC may show an excess of comets with radii 3-6 km \citep{2013Icar..226.1138F}, similar to the above number, and that could account for the diversity of CO abundances within JFCs, although statistics are not sufficient to discern a $Q_{\rm CO}$ vs $D$ trend within the JFC group. While Fig.~\ref{figjfc} is reassuringly consistent with our sublimation calculations for the Kuiper Belt, we note the following
two caveats: (i) the lack of $>$ 5 km JFCs does not allow us to check our prediction that those objects would be less volatile-depleted (ii) the low CO production rate of JFCs may also be related to their repeated perihelion passages on their current orbits. We note finally that, with the notable exception of 29P/Schwassmann–
Wachmann 1, Centaurs, which are dynamically associated with the Scattered Disk and JFC, also appear CO-depleted compared to Oort-cloud comets  \citep[e.g. 10-50 times less CO production per unit surface for Chiron and Echeclus compared to Hale-Bopp;][]{2017AJ....153..230W} in spite of their large $\sim$100
km size. This is a probable consequence of increased outgassing over their $10^6$-$10^7$ years lifetime orbits at giant planet heliocentric distances.

Once CO is released, we find that its dynamics is different from that modelled in extrasolar systems so far (see Appendix \ref{spread}), i.e., the gas does not evolve viscously inwards as expected in massive disks \citep{2016MNRAS.461..845K}. It is due to two reasons. First the gas quantity we find in the KB is very small and not in the fluid regime in contrast to systems detected up to now. Second, the majority of gas detections has been around A-type stars \citep[it is mainly an observational bias because more gas is expected in these systems according to models,][]{2019AJ....157..117M}, where stellar winds are not important (only stars cooler than about F5 possess significant convective envelopes and then magnetic fields that can produce strong stellar winds). In contrast, in the Solar System, the Solar wind (SW) drives the dynamics of the gas. We find that once released, CO gets pushed outwards by SW protons on timescales of a few years (at a rate between $\sim 3$ to 10 au/yr depending on the location, see Appendix \ref{SW}). Some of this CO gets dissociated and ionized (when interacting with SW protons, photons from the Sun and/or the interstellar medium, see Appendix \ref{SW}) on its way out. However, the ionization and dissociation timescales are of order 100 yr so that CO remains the dominant species up to $\sim 500$ au (see Appendix \ref{ion}), i.e. well beyond the heliopause (which is the boundary of the heliosphere where the Solar wind is stopped by the interaction with the local interstellar medium) at $\sim$ 150 au \citep{2020NatAs...4..675O}. The daughter products of the CO dissociation, namely C and O, are ionized in $\sim$ 100 yr, leading to an ionized atomic component beyond $\sim$ 500 au. These ions will then follow the interstellar magnetic lines and get ejected further in the interstellar medium (see Appendix \ref{magn}). The model predictions for CO, C, O (neutral and ionized) as a function of distance to the Sun are shown in Fig.~\ref{fign} and a summary of the model is given in Appendix \ref{gasmod}.

 \begin{figure*}
   \centering
   \includegraphics[width=14cm]{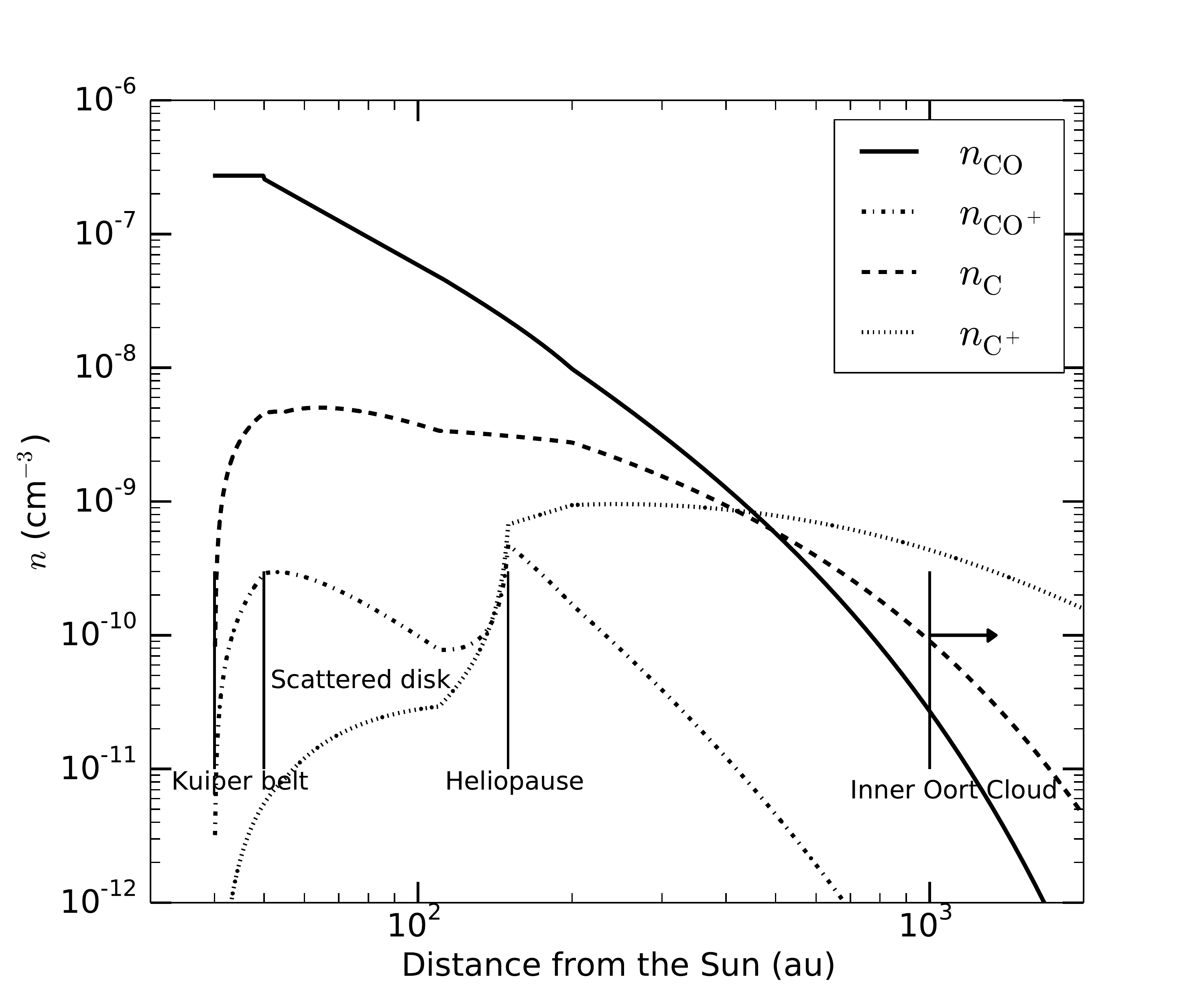}
   \caption{\label{fign} Results of our model for the number density of CO (solid), C and O (dashed), C$^+$ and O$^+$ (dotted) and CO$^+$ (dash-dotted) as a function of distance in our Solar System. The exact shape of the radial profile depends on the heliopause location and geometry, which is not fully modelled in this paper. The assumptions behind this plot are described in the appendix \ref{SW}. Note that the neutral and ionized oxygen number densities are roughly superimposed on those of carbon.}
\end{figure*}

Our model leads to a gas wind with a total CO mass (up to 2000 au) of $\sim 2 \times 10^{-12}$ M$_\oplus$ (i.e., 20 times the CO quantity that was lost by the Hale-Bopp comet during its 1997 passage) and an atomic wind of $\sim 10^{-11}$ M$_\oplus$ as summarized in Table \ref{tab:table2} (and see Appendix \ref{mass}). The CO density in the belt is predicted to be $3 \times 10^{-7}$ cm$^{-3}$ and the column densities along the midplane are of order $\sim 10^8$ cm$^{-2}$ for CO, C, and O. We note that our predictions for O may increase by a factor a few if planetesimals in the KB routinely include O$_2$ ices in quantity similar to CO \citep[as may be expected from recent in situ observations of the comet 67 P/Churyumov–Gerasimenko,][]{2015Natur.526..678B} as they are even more volatile than CO.

\begin{table}

  \begin{center}
    \caption{Results from our gas release model for the total masses, number densities in the belt and column densities along the line of sight to a star.}
    \label{tab:table2}
    \begin{tabular}{|l|c|}
    \hline
    Species & value \\ 
            \hline           \hline             
            \multicolumn{2}{|l|}{Total Masses in M$_\oplus$}\\
            
            \hline  \hline

     CO & $2 \times 10^{-12}$  \\ 
     CO$^{+}$ & $2 \times 10^{-14}$\\ 
     C$^0$ & $ 10^{-12}$ \\ 
     C$^+$ & $6 \times 10^{-12}$ \\ 
     O$^0$ & $ 10^{-12}$ \\ 
     O$^+$ & $8 \times 10^{-12}$ \\ 

     \hline           \hline             
            \multicolumn{2}{|l|}{number densities in cm$^{-3}$ at the center of the belt (45 au)}\\
   
            \hline  \hline

     CO & $3 \times 10^{-7}$  \\ 
     CO$^{+}$ & $3 \times 10^{-10}$\\ 
     C$^0$ & $4 \times 10^{-9}$ \\ 
     C$^+$ & $5 \times 10^{-12}$ \\ 
     O$^0$ & $4 \times 10^{-9}$ \\ 
     O$^+$ & $5 \times 10^{-12}$ \\ 

      \hline       \hline
               \multicolumn{2}{|l|}{column densities along the midplane of the belt in cm$^{-2}$}\\
            
            \hline  \hline

     CO & $2 \times 10^{8}$  \\ 
     CO$^+$ & $7 \times 10^{5}$  \\ 
     C$^0$ & $2 \times 10^{7}$ \\ 
     O$^0$ & $2 \times 10^{7}$ \\ 
     C$^+$ & $10^{7}$ \\ 
     O$^+$ & $10^{7}$ \\ 

\hline
  
      \end{tabular}
  \end{center}

\end{table}

In our Solar System, some cometary models predict that planetesimals should still be outgassing in the KB (see Appendix \ref{gasprod}) at a low rate \citep{2008AJ....135..400J} and it was recently validated through observations \citep{2021arXiv210206313J}. Upper limits in CO from sub-mm studies targeting specific KBOs show that ALMA can detect CO outgassing rates of $2 \times 10^{-8}$ M$_\oplus$/Myr for a specific comet at KB distances \citep{2008AJ....135..400J}. However, in the case of the KB, the release is more diffuse as the emission comes from many KBOs and it would be difficult to observe because of the lack of spatial contrast compared to extrasolar systems where most emission comes from a few beams. We find that Planck and ALMA (in a total power array mode) do not have enough sensitivity to detect the CO rotational lines of the diffuse wind (see Appendix \ref{obs}).
The gas accumulated in the midplane of the KB along the line of sight to a background star would create some absorption in the UV on the star spectrum that could be identified as gas in our Solar System. However, we find that only future instruments may be able to detect this faint absorption. The most promising technique would be to use in-situ missions similar to New Horizons to detect emission of resonance line scattering of carbon and/or oxygen excited by the Sun's UV light (see Appendix \ref{obs}). We find that a super-Alice instrument similar to the current Alice UV probe on New Horizons \citep{2008SSRv..140..155S} but with a larger effective area would reach the low column density level predicted for atoms in the KB. Super-Alice could be built with current technology.



As detailed in Appendix \ref{centaur}, we also explored the CO release from Centaurs and find that their current CO mass loss rate for bodies larger than 4 km is of the same order of magnitude as that predicted by our model for the KB. However, Centaurs being closer to the Sun, CO would be blown out by the SW much faster than in the KB, thus reducing the total CO mass or column that could be observed. We note that the spatial and velocity distributions of this gas is very different (much closer in and faster) to that of gas released in the KB, which could allow future observations to distinguish both components.

The presence of current gas predicted by our model in our Solar System would not impact the dynamics of bodies (dust or planetesimals) evolving around the KB. However, we note that in the past, when the Kuiper belt was much younger and heavier, the release of CO would have been orders of magnitude larger (above the solid line in Fig.~\ref{fig1}) and the gas dynamics would have also been much different (e.g., in the fluid/hydrodynamic regime), potentially leading to some interesting effects, such as delivering some CO mass from the KB to planetary atmospheres as proposed recently for extrasolar systems \citep{2020NatAs...4..769K}. Indeed, in more massive disks, gas becomes optically thick to the SW and it does not get pushed outwards. Instead, gas drifts inwards because of viscous evolution and it can end up accreted onto planets (see Appendix \ref{spread}). The initial KB may have been much more massive before potential dynamical instabilities \citep[e.g.,][]{2005Natur.435..466G} and could have led to CO outgassing rate close to the dashed line in Fig.~\ref{fig1}, hence providing CO that falls onto the young planets in greater quantity than other potential sources such as impacts \citep[see comparison between different CO sources in][]{2020NatAs...4..769K}. This is a whole new study that emerges naturally from this work and will be tackled in a different paper.

\section{Conclusions}
We predict the existence of a gas wind in our Solar System starting at the Kuiper belt and extending farther out. Our model shows that large kilometre-sized planetesimals can still lose volatiles after billions of years of evolution due to the slow heating from the Sun, which warms bodies up at greater depths as time goes by. This finding may provide a clue as to why Jupiter-family comets, thought to originate in the Kuiper belt, are deficient in CO compared to Oort-clouds comets. The released CO gas in the Kuiper belt is constantly produced and then pushed away by the Solar wind, establishing a quasi steady-state CO disk close to the belt with a calculated current total CO mass of $\sim 2 \times 10^{-12}$ M$_\oplus$ (i.e., 20 times the CO quantity that was lost by the Hale-Bopp comet during its 1997 passage). We predict a CO density in the belt of $3 \times 10^{-7}$ cm$^{-3}$, as well as, the existence of a slightly more massive atomic gas wind made of carbon and oxygen (neutral and ionized) with a mass of $\sim 10^{-11}$ M$_\oplus$. This gas cannot be observed with current instrumentation but could be observed with future in-situ missions (e.g., a UV instrument similar to Alice/New Horizons but with a larger detector), and may have played an important role for, e.g., planetary atmosphere formation in the Solar System youth when the gas release rate was much higher, i.e., when the Sun was a few tens of Myr old. Lastly, we show that our new model of gas release due to slow heating of planetesimals by stellar radiation is promising to explain gas detected in exoplanetary system, which would provide the first real physical mechanism for the origin of the gas.

\begin{acknowledgements}
This paper is dedicated to Inaya. We thank the referee for a prompt and helpful review. QK thanks Rosine Lallement and Jean-Loup Bertaux for discussions about the Solar Wind, Heliopause, and Local ISM. QK thanks François Lévrier and Clément Walter for providing information about Planck. QK thanks Andrew Shannon for discussions about latest Kuiper belt collisional models. 
\end{acknowledgements}


\begin{appendix}




\section{Sublimation calculations}\label{subl}

There are three important timescales for gas release through sublimation. First, the planetesimals need to heat up (via conduction due to the Solar influx) to above the CO sublimation temperature of $\sim$ 25 K \citep{2006hgdc.conf.....H} on the thermal timescale; Second the transition from solid to gas (sublimation) must happen on the sublimation timescale; and then the gas must make its way up through the planetesimal pores to finally escape the body on the gas diffusion timescale.

The thermal timescale to heat a layer of thickness $\Delta p$ is given by $\tau_{\rm th}=(\Delta p)^2  / K$, where $K=\kappa/(\rho c_p)$ is the thermal diffusivity (in m$^2$/s), which we will assume to be of order $10^{-10}$ for comet-like objects \citep{2004come.book..359P}, with $\rho$ the planetesimal bulk density in kg/m$^3$, $c_p$ its specific heat in J/kg/K and $\kappa$ its thermal conductivity in J/m/s/K. We note that the thermal diffusivity is smaller for cometary material compared to solid amorphous water ice because cometary material is a porous mixture of ices and refractories including organics. The effects of porosity on the actual effective thermal conductivity (hence diffusivity) are consequential \citep[e.g.,][]{2016A&A...588A.133F,2019A&A...630A...5H}. This value is consistent with current measurements of the thermal inertia at the surface of comets \citep[see][for a review]{2019SSRv..215...29G}. We calculate that during the Solar System lifetime, i.e. $t_S=4.6$ Gyr, the depth to which planetesimals can heat up to the equilibrium temperature of $\sim$40 K \citep{2018ApJ...864...78K} is $\sqrt{t_S K} \sim 3.8$ km. After 4.6 Gyr, the layers deeper than 4 km should still retain their primordial temperature of 10-20 K \citep{2006hgdc.conf.....H,2018ApJ...864...78K} and planetesimals smaller than about 4 km should have no further gas to release as any primordial CO would have already been lost.

The sublimation timescale $\tau_{\rm sub}$ is given by $\rho/(S P_{\rm CO} \sqrt{m_{\rm CO}/(2 \pi k_b T)})$, where $S=3 (1-\Psi)/r_p$ is the total interstitial surface area of the pores \citep[of radius $r_p$ of order 1 micron,][]{2004come.book..359P} of the material per given bulk volume (with $\Psi$ the porosity taken to be 0.6), $P_{\rm CO}= A_{\rm CO} \exp(-B_{\rm CO}/T)$ is the CO saturated vapour pressure  \citep[with $A_{\rm CO}=0.12631$ in $10^{10}$ Pa, and $B_{\rm CO}=764.16$ in K,][]{2004come.book..359P}, and $m_{\rm CO}$ is the mass of a CO molecule. For the temperatures and pressures involved, we find that it takes some $10^3$ yr for a solid CO molecule to turn into its gaseous form.


The gas diffusion timescale is given by $\tau_{\rm dif}=3/4 (\Delta p)^2 (2 \pi m_{\rm CO}/(k_bT))^{0.5} /(\Psi r_p)$. For the temperatures involved, we find that it takes $10^4$ yr to diffuse upwards from 4 km deep.

The time to heat up the planetesimals significantly is longer by orders of magnitude compared to the time to sublimate or diffuse up. Hence, the thermal timescale will set the gas release rate in planetesimals. Let us model the gas release rate due to thermal heating over time.

First, we assume that the CO mass contained in $N_b$ bodies of size $s$ \citep[taken from a state-of-the-art collisional model of the KB,][]{2021Icar..35614256M} within a layer $\Delta p=\sqrt{K t}$ deep is $M_{\rm CO}=4/3 \pi \rho f_{\rm CO} N_b (s^3-(s-\Delta p)^3)$, where $f_{\rm CO}$ is the CO to solid mass fraction (around $f_{\rm COinit}=10\%$ in comets).

The derivative of the CO mass that is warmed up by the Sun is ${\rm d} M_{\rm CO}/{\rm d}t=2 \pi \rho N_b f_{\rm CO} (s-\sqrt{Kt})^2 \sqrt{K/t}$. To compute $f_{\rm CO}$ for different sizes, we calculate for each size bin and at each timestep $\Delta t$ the potential CO mass $M_{\rm COinit}$ contained in the $\Delta p$ layer (assuming nothing was lost) as well as the CO mass that was already lost at time $t$, $M_{\rm lost}=\sum_t {\rm d} M_{\rm CO}/{\rm d}t  \, \Delta t$, and we get $f_{\rm CO}=f_{\rm COinit} (1-M_{\rm lost}/M_{\rm COinit})$, yielding in turn ${\rm d} M_{\rm CO}/{\rm d}t$. This model reproduces the expectation that after 4.6 Gyr of thermal evolution, it is only planetesimals bigger than 4 km that can participate in the gas release as smaller bodies have lost all their CO by that time. The decrease of the gas release rate with time is mostly due to having less and less bodies that can participate in releasing CO. The 10-50 km bodies dominate the gas release in this model. The resulting CO production rate is shown in Fig.~\ref{fig1}. A single 30 km radius planetesimal would currently release around $10^{-14}$ M$_\oplus$/Myr of CO in this model, i.e. much lower than what can be detected with missions targeting specific KBOs \citep{2021Icar..35614072L}.

Scattered disk objects are replenished from various sub-populations of the Kuiper Belt, and possibly the Oort Cloud. Their dynamical lifetime is rather long \citep[$\sim$1.8 Gyr,][though smaller than the age of the Solar System by a factor 2.5]{gome08} but the thermal effect on CO sublimation during this period is limited. Objects in the Scattered Disk spend most of their time at heliocentric distances larger than in the Kuiper Belt. The time spent close to perihelion is limited, so the layer heated by such passages (10 m at most, computed from the orbital skin depth, see Prialnik et al. 2004) remains much smaller than the 4 km where the CO sublimation front is located (after thermal evolution in the Kuiper Belt). This means that our simple thermal model, which is only looking at the deepest layer that can release CO (that only depends on the material as it is a diffusion calculation) will not be affected. We also note that an SD object will spend most of its time in the KB before going to the SD and finally be ejected so that the SD phase is not dominant overall.

The equilibrium temperature in the Oort Cloud as computed through the same energy balance at the surface is extremely low. Other processes may increase it \citep[e.g., cosmic rays etc, as described for example by][]{2021JGRE..12606807D} but the expected equilibrium temperature is roughly 6-10K \citep{jew04}, i.e., well below the sublimation temperature of CO. Therefore, our model does not predict sublimation of CO for Oort cloud objects.

We note that in our model, the CO released in the KB would be coming from pure CO ices and not CO trapped in water, CO$_2$, or other less volatile ices \citep{kouchi}, which sublimation temperatures are too high to become gas at KB distances. Therefore, we only expect the most volatile species to be able to be released as gas in the KB. The volatiles with sublimation temperatures lower than 40 K are N$_2$ (22 K), O$_2$ (24 K), CO (25 K), and CH$_4$ (31 K) with the sublimation temperature given in parenthesis \citep{1985A&A...142...31Y}. The photodissociation timescales can also affect the observability of these species. At the KB, we find photodissociation timescales of 25-62 yr for N$_2$, 8-13 yr for O$_2$, 34-86 yr for CO and 3-8 yr for CH$_4$ \citep[where the min/max values correspond to the Sun at its maximum/minimum activity and the mean over a Solar cycle of 11 yr should be closer to the longer photodissociation timescale,][]{2015P&SS..106...11H}. Now, we use the cometary abundances of volatiles as a proxy for the abundance of planetesimals in the KB. N$_2$ has been observed in a few comets from the ground \citep[e.g., for the comet C/2016 R2 (PanSTARRS),][]{2019A&A...624A..64O} and in-situ in 67 P \citep{2015Sci...348..232R}. At most the N$_2$-to-CO ratio could be 0.06 \citep[][but it is one order of magnitude lower in 67 P]{2019A&A...624A..64O} in comets formed at large distances. N$_2$ would thus be even more difficult than CO to observe even though they have similar photodissociation timescales. O$_2$ has also been observed in the coma of 67 P showing that it may be as abundant as CO on the comet \citep{2015Natur.526..678B}. O$_2$ photodissociates roughly 4 times as fast as CO at the KB and the daughter neutral oxygen species would accumulate to that created from CO if planetesimals in the KB routinely include O$_2$ ices. As for methane, the CH$_4$-to-CO ratio is roughly 0.1 \citep{2017RSPTA.37560252B} and its photodissociation timescale at the KB is roughly 11 times smaller than for CO. Therefore, the released CH$_4$ would also be more difficult to observe than CO.

We also note that collisions would increase the predicted rate because they would expose some fresh CO ices that can be released faster than on a thermal timescale but the collisional timescale for bodies larger than 4 km are longer than the age of our Solar System and this contribution will be negligible for the current KB. Collisions that happened in the early stages of the Trans-Neptunian disk could have affected the size distribution of small bodies and release CO on the surface of these bodies but we note that we use a state-of-the-art size distribution based on observations \citep{2021Icar..35614256M}, which already accounts for previous evolution. Therefore the early collisional evolution is implicitly taken into account in our calculations.

\begin{table}

  \begin{center}
    \caption{Data of JFCs, HFCs, OCCs and Centaurs for which we both have measurements of $Q_{\rm CO}$ and diameter. $Q_{\rm CO}$ is the CO production rate in $s^{-1}$ and its error bar is the 1$\sigma$ uncertainty, $D$ is the equivalent nucleus diameter in km and $r_h$ the heliocentric distance of the object in au. The upper limits are represented by the 3$\sigma$ uncertainties.}
    \label{tabcom}
    \begin{tabular}{|l|c|c|c|c|}
    \hline
    Name & $Q_{\rm CO}$ & $D$ & $r_h$ & ref \\ 
            
            \hline

      C/1995O1& $6.5\pm0.8 \times 10^{27}$ & 69.9 & 14.07 & (1) \\ 
      C/1995O1 & $2.15\pm0.17 \times 10^{30}$ & 69.9 & 0.917 & (1) \\ 
      C/1996B2 & $1.2\pm0.3 \times 10^{28}$ & 2.29 & 1.852 & (2) \\ 
      C/1996B2 & $10.4\pm2.4 \times 10^{28}$ & 2.29 & 0.478 & (2) \\ 
      C/2020F3 & $1.3\pm0.5 \times 10^{28}$ & 5.0 & 0.71 & (3) \\ 
      10199 & $<15.0 \times 10^{27}$ & 249 & 13.5 & (4,5) \\ 
      1P & $4.5\pm2 \times 10^{28}$ & 9.3 & 0.92 & (6)  \\ 
      2P & $1.8\pm0.2 \times 10^{26}$ & 7.02 & 0.473 & (7) \\ 
      8P & $2.4\pm0.4 \times 10^{26}$ & 4.93 & 1.027 & (8) \\ 
      9P & $8.7\pm1.6 \times 10^{26}$ & 6.07 & 1.506 & (8) \\ 
      10P & $<7.8 \times 10^{27}$ & 9.24 & 1.482 & (9) \\ 
      19P & $<4.5 \times 10^{27}$ & 4.34 & 1.360 & (10) \\ 
      21P & $4.6\pm0.8 \times 10^{26}$ & 3.6 & 1.18 & (11) \\ 
      21P & $2.5\pm0.3 \times 10^{26}$ & 3.6 & 1.10 & (11) \\ 
      22P & $<8.1 \times 10^{25}$ & 3.77 & 1.61 & (12) \\ 
      29P & $3.5\pm1.5 \times 10^{28}$ & 56.3 & 6.0 & (13,14) \\ 
      41P & $<8.4 \times 10^{26}$ & 1.4 & 1.06 & (15) \\ 
      45P & $1.9\pm0.3 \times 10^{26}$ & 0.92 & 0.555 & (16) \\ 
      46P & $<3.3 \times 10^{25}$ & 1.2 & 1.13 & (17) \\ 
      67P & $3.4\pm2.5 \times 10^{26}$ & 3.81 & 1.28 & (18,19) \\ 
      73PC & $7\pm2 \times 10^{25}$ & 1.61 & 0.950 & (8) \\ 
      81P & $>5.3 \times 10^{25}$ & 4.17 & 1.74 & (12) \\ 
      95P & $1.3\pm0.6 \times 10^{28}$ & 155 & 8.5 & (20,21) \\ 
      95P & $<8.1 \times 10^{27}$ & 155 & 8.48 & (22,21) \\ 
      95P & $<5.1 \times 10^{28}$ & 155 & 8.9 & (4,21) \\ 
      103P & $2.6\pm0.3 \times 10^{25}$ & 0.68 & 1.064 & (23) \\ 
      174P & $8.0\pm3.0 \times 10^{26}$ & 64 & 6.13 & (14) \\

\hline
  
      \end{tabular}
           {\raggedright References: (1): \citet{2002EM&P...90....5B}; (2): \citet{1999AJ....118.1850B}; (3): Biver al. (in prep) (4): \citet{2001A&A...377..343B}; (5): \citet{2021A&A...652A.141M}; (6): \citet{1997ApJ...475..829F}; (7): \citet{2018AJ....156..251R}; (8): \citet[][and references therein]{2016Icar..278..301D}; (9): \citet{2012A&A...539A..68B}; (10): \citet{2004Icar..167..113B}; (11): \citet{2020AJ....159...42R}; (12): \citet{2012ApJ...752...15O}; (13): \citet{1995Icar..115..213C};  (14): \citet{2017AJ....153..230W}; (15): \citet{2021A&A...651A..25B}; (16): \citet{2017AJ....154..246D}; (17): \citet{2021PSJ.....2...21M}; (18): \citet{2019A&A...630A..19B}; (19): \citet{2020MNRAS.498.3995L}; (20): \citet{2017PASP..129c1001W}; (21): \citet{2017A&A...608A..45L}; (22): \citet{1997P&SS...45..799R}; (23): \citet{2011ApJ...734L...5W}; \par}

  \end{center}

\end{table}

\section{Gas production rate calculations}\label{gasprod}

We use two different techniques to estimate the gas production rate in the KB. First, we assume that the model of gas production that fits detections and non-detections of gas in extrasolar systems \citep{2017MNRAS.469..521K,2019MNRAS.489.3670K} is valid for the KB (Model 1). Second, we test a more physically motivated model that works out the sublimation rates of the bodies in the KB to then derive the total gas production rate in the belt (Model 2).

Model 1 states that the gas production rate is proportional to the mass loss rate of the planetesimal belt collisional cascade. This is because it is assumed that gas is produced owing to collisions when solids grind down to dust somewhere along the cascade. The model does not say which solid size bodies release gas and what is the physics behind it. Rather the model assumes that all CO contained on a large body ($\sim$10\% of its mass) is released before it is ground down to dust and ejected because of radiation pressure. The gas release physics is not yet known and it could be due to high-velocity collisions at the bottom of the cascade, to photodesorption or to sublimation (which this paper may favour), which are all more active for more massive belts releasing more dust indeed. Therefore, the CO gas production rate is roughly equal to the CO fraction of planetesimals times the dust mass loss rate \citep[which is the rate at which mass is passed down the cascade from one bin to the other, which is constant throughout the cascade as the new mass injected at the top of the cascade is lost at the bottom of it at steady-state,][]{2011CeMDA.111....1W}.

We first compute the mass loss rate from a state-of-the-art model of the KB \citep{2012A&A...540A..30V}. Using the cross-section density per size decade $A$ (in m$^2$/m$^3$) derived from their simulations \citep{2012A&A...540A..30V}, we can compute the total mass of bodies in a disk of area $S_{\rm KB}=2 \pi R_{\rm KB} \Delta R_{\rm KB}$ and scale height $H \sim 0.4 R_{\rm KB}$ \citep{2002ARA&A..40...63L} as $M_d(s)=A(4/3 \pi s^3 \rho) S_{\rm KB} (2 H) /(\pi s^2)$ in a given size bin $s$. For the KB location $R_{\rm KB}$ and its width $\Delta R_{\rm KB}$, we take 45 and 10 au, respectively \citep{2002ARA&A..40...63L}. Then, we derive the dust mass loss rate as being $\dot{M}_d(s)=M_d(s)/t_{\rm surv}(s)$, where $t_{\rm surv}(s)$ is the lifetime of a solid of size $s$ taken from collisional simulations \citep{2012A&A...540A..30V}. Assuming a constant CO fraction $f_{\rm CO}$ of 0.1 on solid bodies, we then derive the CO gas production rate as being $\dot{M}_{\rm COtot}=f_{\rm CO} \dot{M}_d(s)$, which is constant at all sizes $s$ (because at steady-state the rate of solids that are broken up by collisions between large bodies is equal to the dust mass loss rate due to radiation pressure). Indeed, making the calculation using the small micron-sized dust grains at the bottom of the cascade or for larger bodies at collisional equilibrium, we find $\dot{M}_{\rm COtot}=10^{-9}$ M$_\oplus$/Myr.

We also derive the mass loss rate based on measurements of the student dust counter on the New Horizons mission \citep{2008SSRv..140..387H}. The number of particles between 0.5 and 5 microns hitting the student dust counter in the KB at 45 au, moving at $v_{\rm NH} \sim 14$ km/s, is estimated to be $F_d=1.5 \times 10^{-4}$ m$^{-2}$ s$^{-1}$ \citep{2019ApJ...881L..12P}. The total mass of grains is then given by $M_d[0.5-5{\mu}m]=(F_d / v_{\rm NH}) \, m^* S_{\rm KB} (2 H)$, where $m^*$ is the mean mass of a particle in the 0.5-5 micron size range given a particle size distribution slope in -2.5 as it is in the PR-drag (rather than collisional) regime \citep{2011CeMDA.111....1W,2012A&A...540A..30V}. To get the mass loss rate of grains between 0.5 and 5 microns, we divide by the PR-drag timescale $t_{\rm PR}=400 \beta^{-1} (R_{\rm KB})^2$ \citep[in yr,][]{2005A&A...433.1007W}, where we take the ratio between the radiation pressure force to that of stellar gravity $\beta=0.2$ based on the mean size of a grain of mass $m^*$ \citep{2012A&A...540A..30V}. However, to get the mass loss rate of the cascade (and not just that of the grains captured by the student dust counter), we need to multiply by the number of logarithmic bins up to the size at which collisions dominate over PR-drag \citep{2011CeMDA.111....1W}, i.e., up to $s_{\rm pr}=100$ microns \citep{2012A&A...540A..30V}. We obtain $\dot{M}_{\rm COtot}=9 \times 10^{-10}$ M$_\oplus$/Myr, i.e. a gas production rate very close to that found with the previous method.

Model 2 uses the sublimation model described in the previous section. We find that the gas sublimation rate for the KB is dominated by large bodies $>4$ km after 4.6 Gyr evolution. Below this size, all CO gas was released because the entire CO inventory has been sublimed already and the gas production rate drops to zero. To get the final CO production rate, we sum over all the size bins and find $2 \times 10^{-8}$ M$_\oplus$/Myr, which is roughly a factor 20 higher than the previous estimates. The temporal evolution is shown in Fig.~\ref{fig1}.

We note that comet sublimation models still predict outgassing at large KB-like distances and they estimate that a single large Hale-Bopp comet at 40 au would release about $10^{-10}$ M$_\oplus$/Myr of CO \citep{2008AJ....135..400J}. However, this is valid only if there remains CO ice on the planetesimal surface, but as we showed previously it would be long gone from the upper layers after 4.6 Gyr evolution, and our model predicts a rate about 4 orders of magnitude smaller for a given large planetesimal similar to Hale-Bopp.

We also check whether the sublimation model is consistent with the high release rates observed around younger more massive stars with ALMA. Let us take an extreme case, which is that of a belt as massive as that of the $\beta$ Pic system \citep[i.e., more massive than 1000 M$_\oplus$ if we assume the belt to be composed of bodies up to 100 km, though bodies may be born smaller,][]{2021MNRAS.500..718K}. To reach a belt of 1000 M$_\oplus$, we scale up the number of bodies in the KB (of total mass equal to $\sim 0.1$ M$_\oplus$) by a factor $10^4$ and re-run our model leading to the dashed line in Fig.~\ref{fig1}. This shows that gas release rates of $\sim 0.1$ M$_\oplus$/Myr can be reached with this model when gas is initially released from the young belt. 

The gas release rate in the belt of the $\beta$ Pic system could be up to 0.1 M$_\oplus$/Myr \citep{2016MNRAS.461..845K} but we note that the gas release rate necessary to explain the CO mass observed could be lower, as there could be sufficient carbon to shield CO from photodissociation in this system \citep{2019MNRAS.489.3670K}. The temperature conditions of planetesimals in the KB could be similar to that of other exo-Kuiper belts, as belts tend to form preferentially at a given distance from their star \citep[$R \propto L^{0.19}_\star$,][]{2018ApJ...859...72M}, and so the resulting belt temperature ($\propto L_\star^{0.16}$) is only weakly dependent on the stellar luminosity (56 K for the $\beta$ Pic extrapolation) and close to that of the KB.
We note that the size distribution in young systems could be different to that of the Solar System, which could also increase/decrease the model predictions but this is not taken into account here. A more thorough study (including collisions for young systems) of this model (and its free parameters such as the material composition or temperature of the belt) and whether it can explain all observations (or just that of the less massive belts for instance) is beyond the scope of this KB-focused paper.

\section{Description of the gas evolution model}\label{gasmod}

We now summarize the model for the evolution of the gas released in the KB so that the reader gets a feel for what mechanisms are at play while reading the more in-depth sections that follow.

The main ingredients used in the gas evolution model for the KB are summarized in Table~\ref{tab:table1} and the main timescales in Table~\ref{tab:tablesum}. The main ideas go as follows: 1) CO is released from planetesimals in the KB. 2) CO is quickly pushed outwards with a velocity $\gtrsim$ 3 au/yr due to collisions with high-velocity ($\sim$ 400 km/s) protons from the solar wind (and a small fraction of the CO gets ionized and dissociated during these collisions and due to impinging photons from the Sun and the interstellar medium). 3) Most CO gets pushed beyond the heliopause \citep[located at $\sim$150 au,][]{2017NatAs...1E.115D,2019NatAs...3.1019R} before it has time to dissociate or ionize. 4) CO finally turns into C+O and C gets ionized due to photons from the interstellar medium, while O gets ionized due to collisions with protons from the local cloud of the interstellar medium that is colliding with our Solar System. 5) Then the ionized atoms follow the interstellar magnetic field lines and get ejected further into the local interstellar medium. 
More details about each step of the model are given in the following sections.

\begin{table*}
  \begin{center}

  \begin{threeparttable}

    \caption{Processes accounted for in this study for the CO evolution once released from the Kuiper belt. We show the dominant process at the Kuiper belt location in the column with a * symbol.}
    \label{tab:table1}
    \begin{tabular}{|l|l|c|c|c|}
    \hline
    Processes & interactions & quantity of interest & value & *\\ 
            \hline           \hline             
            \multicolumn{5}{|l|}{CO}\\
            
            \hline  \hline

      \multirow{2}{*}{Ionization} & SW protons & charge exchange cross section \citep{2017PhPro..90..391L} & $1.5 \times 10^{-15}$ cm$^{2}$ &  \\ 
      & solar photons &  photoionization rate (at 1 au) \text{\citep{1992Ap&SS.195....1H}} & $6 \times 10^{-7}$ s$^{-1}$ & x\\ 
      & SW e$^{-}$ impacts&  ionization rate \citep{2009Icar..199..505R} & $5.49 \times 10^{-9}$ cm$^3$/s  & \\ 
      \hline       \hline
      \multirow{2}{*}{Dissociation} & SW protons & dissociation cross section \citep{2017PhPro..90..391L} & $1.5 \times 10^{-17}$ cm$^{2}$  & \\ 
      & solar photons  & photodissociation rate (at 1 au) \text{\citep{2015P&SS..106...11H}} & $5 \times 10^{-7}$ s$^{-1}$  & x\\ 
      & ISM photons & photodissociation rate \text{\citep{2017A&A...602A.105H}} & $2.4 \times 10^{-10}$ s$^{-1}$  & \\ 
      \hline
      \multirow{2}{*}{Collisions} & SW protons & radial velocity after first collision & $\sim$ 3 au/yr & x\\ 
      &  & collisional frequency (at 45 au) & $\sim$ 2.7 yr & \\
      & LISM protons & radial velocity after collision  & depends on $v_{\rm{CO}}$ & \\ 
      & & collisional frequency & $\sim$ 1.7 yr & \\

      \hline    
      \hline                  
            \multicolumn{5}{|l|}{CO$^+$}\\
            
            \hline  \hline

           \multirow{2}{*}{Dissociation} & solar photons  & photodissociation rate (at 1 au) \text{\citep{2017A&A...602A.105H}} & $5 \times 10^{-8}$ s$^{-1}$  & \\ 
      
      & ISM photons & photodissociation rate \text{\citep{2017A&A...602A.105H}} & $1 \times 10^{-10}$ s$^{-1}$ & x \\ 
      \hline
      \multirow{2}{*}{Collisions} & SW protons & radial velocity after first collision & $\sim$ 3 au/yr & x\\ 
      &  & collisional frequency (at 45 au) & $\sim$ 24 min & \\
      & LISM protons & radial velocity after collision  & depends on $v_{\rm{CO+}}$ & \\ 
      & & collisional frequency & $\sim$ 15 min & \\

      \hline    
      \hline                       
            \multicolumn{5}{|l|}{C}\\
            
            \hline  \hline

      \multirow{2}{*}{Ionization} & solar photons &  photoionization rate (at 1 au)  \text{\citep{2015P&SS..106...11H}} & $4 \times 10^{-7}$ s$^{-1}$ &  \\ 
      & ISM photons &  photoionization rate \text{\citep{2017A&A...602A.105H}} & $3.4 \times 10^{-10}$ s$^{-1}$  & x\\ 
          \hline
      \hline
         
            \multicolumn{5}{|l|}{O}\\
            
            \hline  \hline

      \multirow{2}{*}{Ionization} & SW protons & charge exchange cross section$^a$ \text{\citep{1997A&A...317..193I}} & $1.1 \times 10^{-15}$ cm$^2$  & x  \\ 
      & solar photons &  photoionization rate (at 1 au) \text{\citep{2015P&SS..106...11H}}  & $4 \times 10^{-7}$ s$^{-1}$ & \\ 
      & ISM photons &  photoionization rate & $0$ & \\ 
      \hline
      \end{tabular}

\begin{tablenotes}
      \small
      \item $^a$ We note that some extra ionization comes from electron impact ionization while crossing the heliosphere but CO mainly photodissociates further out \citep{1999A&A...344..317I}.
    \end{tablenotes}
  \end{threeparttable}
    \end{center}

\end{table*}

\begin{table*}

  \begin{center}
    \caption{Timescales of the dominant processes at play}
    \label{tab:tablesum}
    \begin{tabular}{|c|c|c|}
    \hline
    Processes & Interactions & Timescale \\ 
            
            \hline           
            \hline     
                    
            \multicolumn{3}{|l|}{CO}\\
            
            \hline  \hline

       Ionization & solar photons \text{\citep{1992Ap&SS.195....1H}} & 107 yr (at 45 au, $\propto 1/r^2$) \\ 
      \hline       \hline
      Dissociation & solar photons \text{\citep{2015P&SS..106...11H}}  & 50 yr (at 45 au, $\propto 1/r^2$), 120 yr (at $>$70 au) \\ 
      \hline
      Collisions & SW protons & 2.7 yr\\ 
      Collisions &  LISM protons & 1.7 yr\\ 

      \hline    
      \hline                  
            \multicolumn{3}{|l|}{CO$^+$}\\
            
            \hline  \hline
      
      Dissociation & ISM photons \text{\citep{2017A&A...602A.105H}} & 305 yr (at $>$40 au)  \\ 
      \hline
      Collisions & SW protons & 24 min \\ 
      Collisions & LISM protons & 15 min \\ 

      \hline    
      \hline                       
            \multicolumn{3}{|l|}{C}\\
            
            \hline  \hline

      Ionization &ISM photons \text{\citep{2017A&A...602A.105H}} & 94 yr (at $>$40 au)   \\ 
          \hline
      \hline
         
            \multicolumn{3}{|l|}{O}\\
            
            \hline  \hline

     Ionization & SW protons \text{\citep{1997A&A...317..193I}} & 160 yr (at 45 au, $\propto 1/r^2$), 110 yr (at $>$150 au)   \\ 
            \hline
      \end{tabular}
  \end{center}

\end{table*}


\section{Ionization fraction for the gas in the Kuiper belt}\label{ion}

We compute the ionization fraction of the main species we study by equating the ionization rate to the recombination rate. For the Solar System, we take the photoionization probabilities at 1 au \citep{2015P&SS..106...11H} (in s$^{-1}$) for C, O, and CO: $8\times 10^{-7}, 4 \times 10^{-7}, 6 \times 10^{-7}$. The recombination timescales are based on the modified Arrhenius equation (in cm$^3$/s) of the form $\alpha (T/300 \, {\rm K})^\beta \exp(-\gamma/T)$. The recombination rate for O$^+$ is given by \citet{1999ApJS..120..131N}, $\alpha=3.24 \times 10^{-12}$, $\beta=-0.66$, $\gamma=0$, for C$^+$ we use \citet{1997ApJS..111..339N}, $\alpha=2.36 \times 10^{-12}$, $\beta=-0.29$, $\gamma=-17.6$, and for CO$^+$ $\alpha=2.75 \times 10^{-7}$, $\beta=-0.55$, $\gamma=0$ are taken from the KIDA database \citep{2012ApJS..199...21W}. 

One striking difference with previous work on the subject of gas in planetary systems (where models were developed mostly for A stars in young systems) is that in the Solar System, oxygen can be ionized because of the presence of numerous UV photons at energies greater than the ionization potential of oxygen (13.6 eV). Solving for the ionization fractions of CO, C and O analytically, we search for the electron density (expected to be the main collider here) necessary to get an ionization fraction greater than 0.5 so that we can later estimate (when we get the electron density from the model) whether the different species will be ionized or not. For CO, we find $n_e<3 \times 10^{-4} (T/30 K)^{0.58}$ cm$^{-3}$. For C \citep[where we also account for the ionization rate of $3.39 \times 10^{-10}$ s$^{-1}$ from the interstellar medium, which is greater than the photoionization rate at 45 au,][]{2017A&A...602A.105H}, we find $n_e<91 (T/30 K)^{0.29}$ cm$^{-3}$ and for O, we have $n_e<13 (T/30 K)^{0.66}$ cm$^{-3}$.

We also estimate the CO ionization rate through SW electron impact. The density and temperature at the KB of the fast moving electrons ($\sim$ 610 km/s) are $10^{-3}$ cm$^{-3}$ and $3 \times 10^5$ K (25.9 eV) \citep{1998JGR...10329705M}. The electron impact ionization rate (for an electron energy of 18 eV) is $k_e=5.49 \times 10^{-9}$ cm$^3$/s, which is multiplied by $25.9/18 \sim 1.4$ to account for the highest velocities \citep{2009Icar..199..505R}. Equating $k_e n_{e{\rm SW}}n_{\rm CO}$ to the recombination rate of CO$^+$ described above, we find that an electron density lower than $8 \times 10^{-6}$ cm$^{-3}$ is necessary to lead to an ionization fraction of CO greater than 0.5 \citep[the slow moving electrons contribution is of the same order of magnitude albeit slightly lower,][]{1998JGR...10329705M}. The photoionization is therefore much more efficient, and electron impacts from the SW can be neglected for CO ionization.

In the SW result section, we will use the electron density given by our model to provide an estimate of the ionization fraction of the different species. In broad terms (for $T$ close to 20-50 K), C, O, and CO get $>50$\% ionized at 45 au for $n_e<100$ cm$^{-3}$, $<10$ cm$^{-3}$ and $<3 \times 10^{-4}$ cm$^{-3}$, respectively. So it is most likely that O and C will be close to 100\% ionized at 45 au (if they can be produced and remain at 45 au for long enough, see later), and for CO it depends on the details that we will investigate in the coming sections (which are complicated by impacts from SW protons pushing CO away faster than it can photoionize).

\section{Spreading timescales}\label{spread}

In current and past works about gas in planetary systems, the viscous evolution of a gas disk is often parameterized using an $\alpha$ model \citep{1973A&A....24..337S}, which provides a good description in the fluid regime when the Knudsen number (mean free path over gas scale height) is lower than 1. The value of $\alpha$ sets how fast the gas disk spreads as the viscous timescale $t_{\rm visc}$ is given by $r^2/\nu$ with the viscosity $\nu=\alpha c_s^2/\Omega$, where $c_s$ is the sound speed and $\Omega$ the orbital frequency. A recent theoretical study shows that the magnetorotational instability \citep[MRI,][]{1991ApJ...376..214B} may be able to develop in the debris disk regime and produce large $\alpha$ values (of the order of 0.1) owing to the high ionization fraction of the gas in these systems \citep{2016MNRAS.461.1614K}. Observations seem to favour high $\alpha$ values as well \citep{2016MNRAS.461..845K,2020MNRAS.492.4409M}, though this may depend on the emergence of non-ideal MRI effects (such as ambipolar diffusion) as well as the magnetic field strength \citep{2016MNRAS.461.1614K}. Taking $\alpha$ between $10^{-4}$ and 0.1, we find that the viscous timescale at the KB location can vary from 1.1 Myr to 1.1 Gyr assuming a gas temperature of 30 K and a mean molecular weight of 28 (gas dominated by CO).

However, the gas density around the KB may be too low to be in the fluid regime, and previous considerations used to describe gas in exoplanetary systems do not apply. When gas density is very low and the Knudsen number becomes greater than 1, then the non-fluid viscosity can be evaluated as follows, $\nu_2=\lambda_{\rm mfp} \, c_s$, with $\lambda_{\rm mfp}$ the mean free path of a gas particle equal to $(n_{\rm gas} \sigma_{\rm col})^{-1}$, where $n_{\rm gas}$ is the gas density and $\sigma_{\rm col}$ its collisional cross-section. Since the gas scale height is $c_s/\Omega$, this regime may happen when $n_{\rm gas} < \Omega/(\sigma_{\rm col} c_s)$. Let us consider the most favourable case of collisions between charged particles (with a greater collisional cross-section) and take $\sigma_{\rm col}=\pi R^2_{\rm cc}$, where $R_{\rm cc}$ is the cross-sectional radius for a proton calculated by equating kinetic energy to electrostatic energy so that we get $R_{\rm cc}=e^2/(6 \pi \epsilon_0 k_{\rm b} T_{\rm gas})$, with $e$ the elementary charge and $\epsilon_0$ the vacuum permittivity. Thus, we find that the non-fluid regime appears when $n_{\rm gas} < n_{\rm crit}=3 \times 10^{-6}$ cm$^{-3}$. As for neutral atoms, Coulomb collisions will happen with charged particles. To find the cross-section of the, e.g., C$^+$-O collisions, we account for the fact that the ion induces a dipole on the neutral atom, which gives birth to an electric repulsion \citep{1989A&A...223..304B}. The cross-section for neutral-ion interactions is roughly $10^4$ times smaller than for proton-proton collisions, hence increasing $n_{\rm crit}$ by a factor $10^4$.

Assuming the CO gas production rate of $2 \times 10^{-8}$ M$_\oplus$/Myr derived earlier, we find that the ionized carbon produced from CO photodissociation must remain at 45 au for $>$ 25 yr to reach a gas density greater than $n_{\rm crit}$ and therefore be in the fluid regime. For the case of neutral atoms (e.g. CO, C or O), they should remain for $>$ 0.25 Myr. We will calculate in the coming section how long the CO, C or O can survive at a given position owing to the action of the Solar wind, which we find acts on smaller timescales than viscous spreading in the Solar System. We find that, due to the SW, the gas density cannot increase sufficiently to be in the fluid regime, which is going to drastically change the dynamics of gas compared to previous studies.

\section{Effect of Solar wind on gas}\label{SW}

Let us consider the effect of the Solar wind on the KB gas ring. The Solar wind medium density is $\sim$8 cm$^{-3}$ at the location of the Earth (1 au), which translates in a density of $n_{\rm SW}=4 \times 10^{-3}$ cm$^{-3}$ at 45 au \citep{2019A&A...632A..89H} (assuming a $1/r^2$ drop-off, i.e. constant radial velocity). We can first calculate the time between collisions with the SW for each molecule, which is given by $1/(\sigma_{\rm X-SW} n_{\rm SW} v_{\rm SW})$, where $\sigma_{\rm X-SW}$ is the cross-section of interaction between X and protons from the SW and $v_{\rm SW}$ is the Solar wind velocity of about 400 km/s. We consider that the cross-section of interaction between ionized SW particles and CO is set by the induced dipole between CO and a proton such as described above in the spreading timescale section. We find that the induced dipole cross-section is $\sigma_{\rm CO-SW} \sim 7 \times 10^{-14}$ cm$^{2}$. For CO, we find that there is a collision with a SW proton every 2.7 yr (i.e., before it has time to dissociate or ionize). For an ionized species (e.g., CO$^+$ and proton) with a higher cross section (see previous section), it is only about 24 minutes. Similar calculations in the local cloud of the interstellar medium (using a velocity of 26 km/s and proton density of 0.1 cm$^{-3}$) lead to 1.7 yr and 15 min for neutral and ionized species, respectively \citep{2019NatAs...3.1019R}.

We can estimate the rate of collisions with the SW protons in the KB for a given density of atoms or molecules as $$W_{\rm SW}=V_{\rm KB} n_{\rm X} \sigma_{\rm X-SW} n_{\rm SW} v_{\rm SW},$$ where $V_{\rm KB}$ is the KB volume and $n_{\rm X}$ the density of X.

\subsection{Basics of the model based on former literature knowledge}

Let us start our calculations as if we were only aware of the models developed for extrasolar systems (mostly for young A stars) as it will ease the transition to a Solar System model where the addition of the Solar wind adds another layer to currently used models. 

\begin{figure}
   \centering
   \includegraphics[width=9.5cm]{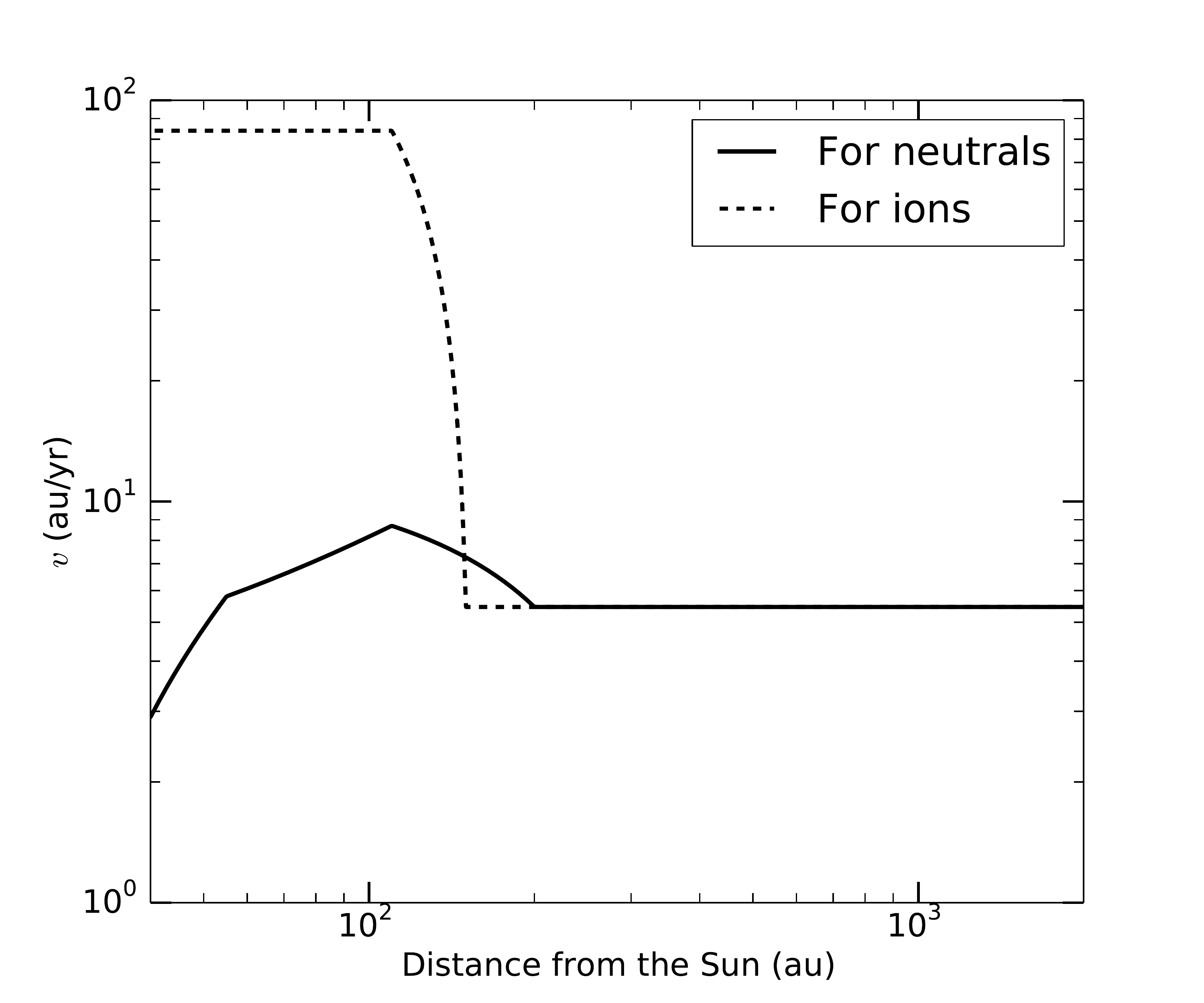}
   \caption{\label{figv} Sketch of the velocity of particles as a function of distance to the Sun for neutrals (solid) and ions (dashed). The exact shape of the radial profile depends on the heliopause location, which is not fully modelled in this paper (and assumed symmetric as recently proposed by new studies). The assumptions behind this plot are described in the appendix section about the effect of SW.}
\end{figure}

First, we consider the effect of the SW on the CO molecules for which we first assume a number density $n_{\rm CO}$ of $5 \times 10^{-6}$ cm$^{-3}$, which is expected if CO can survive for about 50 years against photodissociation (which is close to that value at 45 au in our Solar System). We calculate the mean loss rate of CO due to SW interactions as $\dot{M}_{\rm SW} = W_{\rm SW} \mu m_p$ assuming that after each collision the high impact velocity will give an outwards kick to the CO molecule. Indeed, working out the momenta for the CO molecule and the high-velocity proton and using that it is conserved and that the collision is perfectly inelastic, we find $\vec{p}_{\rm tot}=\vec{p}_{\rm proton}^{\rm \,\, init}+\vec{p}_{\rm CO}^{\rm \,\, init}=\vec{p}_{\rm proton+CO}^{\rm \,\, final}$. Therefore, we can solve for the final CO velocity vector that we find is 13.8 km/s radially and 4.3 km/s azimuthally. It is indeed unbound already after the first kick. We note that if the collision is perfectly elastic, the final radial velocity could be twice as great. In reality, when the CO and the proton stick together, the collision will indeed lead to a radial velocity of 13.8 km/s (2.9 au/yr) and otherwise, it will be a value in the range 13.8-27.6 km/s. Given the charge exchange cross section between SW protons and CO given in Table \ref{tab:table1}, we find that exchanges will happen on a $\sim$100 yr timescale, i.e., the aftermath of the collision is most likely CO but the detailed modelling of the collision geometry and properties goes beyond our simple model. We note that there could be multiple collisions before reaching the heliopause at $\sim$ 150 au. For instance, if the first collision between CO and a proton happens at 45 au after 2.7 yr then the CO will travel radially for a bit longer than 2.7 yr (as the timescale scales as $(r/45)^2$) and we calculate (using a mean collision time over the distance travelled) that the next collision will happen at $\sim$ 55 au and so it will be outside the main KB and its radial velocity will have doubled. The next collision will happen $\sim$ 10 yrs later when the particle is at $\sim$ 110 au, close to the heliopause. After that there are no more collisions with protons from the SW as the SW proton density becomes too small given the fast velocity CO moves at (8.7 au/yr or 41 km/s). Next collisions will be with the protons from the local interstellar medium, whose density is around 0.1 cm$^{-3}$ \citep{1997A&A...317..193I}, finally equalizing the velocities to around 26 km/s after a few years. For ions, the collisions with protons (be they from the SW or interstellar medium) are very fast and it takes a couple of tens of minutes to reach 400 km/s before the heliopause or 26 km/s beyond it. The transitions are smoother than described above and the variation of the velocity with distance is represented in Fig.~\ref{figv}.

Thus, the CO mass loss rate due to the SW (when it can accumulate for the photodissociation timescale of 50 years due to the Sun radiation) would be $\dot{M}_{\rm SW} \sim 4 \times 10^{-7}$ M$_\oplus$/Myr (where $m_p$ is the proton mass and we take $\mu=28$), which is higher than the CO gas production rate we found in this study so that CO will actually be removed before it has time to photodissociate, which will reduce the total CO density we used so far. 

\subsection{Improvement and further complexities of the model}

To find the number density of CO given its production rate and loss through SW interactions, we equate $\dot{M}_{\rm SW}$ and $\dot{M}_{\rm COtot}$ and find $n_{\rm CO} = 3 \times 10^{-7}$ cm$^{-3}$, which is more than one order of magnitude smaller than the value previously found without properly accounting for the dynamical effect of the SW. This is because indeed, there is a collision every 2.7 years at 45 au with high velocity protons so that CO gets removed roughly 20 times faster than under the action of photodissociation alone. CO will move outwards before being eventually photodissociated by UV photons from the interstellar radiation field after roughly 120 yr, i.e., at a distance roughly ten times that of the KB. However, once in the local interstellar medium \citep[i.e., beyond the heliopause at $\sim$ 150 au,][]{2017NatAs...1E.115D}, the velocity of CO molecules will be slowed down by collisions with the local interstellar medium protons every few years.

From the CO number density calculation, it is also clear that the CO will be fully ionized by the UV photons from the Sun because we estimate that the electron density is $<3 \times 10^{-4}$ cm$^{-3}$ assuming that electrons come from CO ionization (see CO$^+$ densities in Fig.~\ref{fign}). However it is not instantaneous and it will take approximately 100 yr to photoionize a given molecule of CO at 45 au \citep{2015P&SS..106...11H}. Also, there will be some charge exchanges between CO and protons from the SW leading to CO$^+$ that may happen faster than photoionization. The cross section of exchange for protons with an energy of $\sim 1$ keV is $\sim 1.5 \times 10^{-15}$ cm$^2$ so that exchanges with a CO molecule will happen every 134 yr, at a slightly lower rate than photoionization. We note that photoionization leading to C$^+$ happens 13 times less often and 15 times for O$^+$ \citep{2009Icar..199..505R}. As for the collisions with protons from the SW, they lead to C$^+$ 5 times less often and to O$^+$ 10 times less often.

As soon as CO becomes CO$^+$, it will leave the system very quickly as collisions with the Solar wind protons happen every 24 min rather than 2.7 yr and the velocity becomes very quickly equal to that of SW protons (i.e., $\sim$400 km/s). But CO will start heading outwards after 2.7 yr anyway and may become CO$^+$ on its way out.
The final quantity of CO in the KB is therefore indeed set by the frequency of impacts with the SW of 2.7 yr. But we note that the escape of CO outside of the KB is not instantaneous as the molecule will travel radially at roughly 14 km/s (or $\sim 2.9$ au per yr) after an impact with a SW proton, and it will take a gas particle in the middle of the KB (45 au) roughly 1.7 yr to reach 50 au, so that CO can slightly accumulate before leaving the belt. Multiplying the CO number density we obtained with $\dot{M}_{\rm SW}=\dot{M}_{\rm COtot}$ by (1.7+2.7)/2.7, we obtain the mean CO density in the belt $n_{\rm CO}$ equal to $5 \times 10^{-7}$ cm$^{-3}$, which will roughly scale as $1/r^2$ until it dissociates or ionizes. This leads to a surface density at 45 au $\Sigma_{\rm CO}=\mu m_p n_{\rm CO} (2 H) \sim 10^{-16}$ kg m$^{-2}$.

\subsection{Modelling of the outer regions of the KB}

We note that CO will photodissociate in $\sim$120 yr beyond 45 au while photoionization will operate on a timescale of 107 ($r$/45 au)$^2$ yr and proton collisions leading to CO$^+$ in 134 ($r$/45 au)$^2$ yr. Because CO moves at a rate of 2.9 au/yr after a collision with a proton, it will be at hundreds of au after 100 yr and the timescales for photoionization and ionization by protons from the SW become $>10,000$ yr. Therefore, CO photodissociates before it has time to be ionized. However, the time for CO to move outside of the KB is on average 4.4 yr and some small fraction of CO will have time to ionize and photodissociate. Using an exponential decay law for the time evolution of ionization and photodissociation, we find that 7.3\% of CO will be ionized  at 45 au and  3.6\% will be photodissociated. If we assume that most electrons in the KB come from CO ionisation then this gives an electron density $n_e$ at least greater than $7.3\% \times n_{\rm CO}=2 \times 10^{-8}$ cm$^{-3}$. CO$^+$ will reach the SW proton velocity in a few hours as it collides every 24 minutes with them. Therefore, we expect the CO$^+$ density in the belt to be $7.3\% \times n_{\rm CO} \times (13.8/400)=2.5 \times 10^{-3} \, n_{\rm CO}=7 \times 10^{-10}$ cm$^{-3}$.


After CO eventually photodissociates, an atomic gas component will appear. This leads to Fig.~\ref{fign} where we show the down-wind profile of gas and assume that the heliosphere is not too asymmetric and close to a ball-shape as recently proposed \citep[][and we do not model the specificities of the heliosheath]{2017NatAs...1E.115D,2020NatAs...4..675O}. We also assume that ions move faster than neutrals as given by Fig.~\ref{figv} to work out the relative gas densities. CO produced in the KB would then cross the heliopause (at $\sim$ 150 au) after roughly 20 yr. Assuming that CO moves at 26 km/s after the heliopause \citep{2020NatAs...4..675O}, CO will be mostly photodissociated (after 120 yr in total) at $\sim$ 500 au (on the downwind side and slightly further in on the upwind side because CO gets pushed backwards once it reaches the heliopause). The carbon and oxygen atoms will eventually ionize. The photoionization timescale for C is 94 yr (owing to the interstellar medium photons). For O it takes $>$13 kyr at $>$400 au from the Sun to become ionized (as only the Sun's photons are energetic enough) but O will cross the heliopause and encounter protons from the local interstellar medium and charge exchanges can then happen that will operate in $\sim$ 110 yr \citep{1997A&A...317..193I}. Therefore, carbon and oxygen will ionize quickly (we assume 100 yr for both in Fig.~\ref{fign}). The ionized carbon and oxygen will start dominating at $\sim500$ au (on the downwind side). They will then follow the interstellar magnetic field lines (see next section) and get ejected further into the local interstellar medium.
 
One of the main conclusions of this model is that CO will move outwards and almost no gas released from the KB will be able to make it inwards towards Neptune. However, we note that it could have been different in the past, as the KB was much more massive and the gas release rate should have been high enough for the gas to be in the fluid regime. In this situation, the gas becomes more optically thick to collisions with protons and may have time to evolve viscously inwards (as described in the viscous evolution section) rather than being pushed outwards, but consideration of this regime is not the purpose of the current paper.

\section{Interaction with the magnetic field}\label{magn}

Let us now analyse the dynamics of an ion produced in the KB choosing CO$^+$ for the example below.

We choose a density $n=10^{-9}$ cm$^{-3}$ of CO$^+$ as a proxy (as it is close to the value we find in the previous section). It leads to a mean free path of $2 \times 10^{17}$ cm for proton-proton collisions (i.e. $>$10,000 au), $3 \times 10^{19}$ cm for charged-neutral collisions and $3 \times 10^{21}$ cm for neutral-neutral species collisions (see SW section for the different cross-sections).
We can then compare these values with the gyroradius $r_g = m v_{\rm kep} / (e B)$. With the interplanetary magnetic field $B \sim 0.1$ nT at the KB \citep[it is 6 nT at Earth and scales as $1/r$,][]{1963ApJ...137.1268A}, we obtain $r_g = 8 \times 10^8$ cm for a molecule of CO.
We can also compare to the relevant lengthscale of the problem $L=R_{\rm KB} (c_s/v_{\rm kep})$, which we find equal to $L \sim 2 \times 10^{13}$ cm.

The gyroradius being smaller than both the mean free path and the relevant lengthscale, we conclude that ionized species produced in the KB will follow the interplanetary magnetic field lines and escape the Solar System.

The same reasoning applies to ionized particles beyond the heliopause in the local interstellar medium, which will then follow the interstellar magnetic lines.

\section{Gas mass, density and column calculations}\label{mass}

Assuming a CO gas production rate of $2 \times 10^{-8}$ M$_\oplus$/Myr and that CO escapes the KB in 4.4 yr (see SW section), we get a total CO mass in the KB (40-50 au) of $\sim 10^{-13}$ M$_\oplus$, which is roughly the total CO mass that was lost by the Hale-Bopp comet in 1997. This mass translates into the previously calculated mean number density of $3 \times 10^{-7}$ cm$^{-3}$ in the KB.
The total CO mass (up to 2000 au) obtained with our model is equal to $2 \times 10^{-12}$ M$_\oplus$ or roughly 20 times the CO mass that was lost by the Hale-Bopp comet during its 1997 passage.


For the carbon and oxygen wind that forms from the CO photodissociated molecules, we find that the total atomic masses (up to 2000 au) are $\sim 6 \times 10^{-12}$ M$_\oplus$ and $\sim 8 \times 10^{-12}$ M$_\oplus$ for neutral and ionized species, respectively. 

For the column densities $N_X$ of species $X$, we integrate along the midplane outwards so that $N_X=\int_{R_{\rm in}}^{R_{\rm out}} n_X \, {\rm d}R$, with $R_{\rm in}=45$ au and $R_{\rm out}=2000$ au, and Table~\ref{tab:table2} summarizes all these calculations.

\section{Comparison to other sources of CO in the Solar System}\label{centaur}

Comets release abundant CO when they approach the Sun \citep[e.g.][]{2004come.book..391B}. For instance, the 60 km diameter Hale-Bopp comet released a CO mass of $\sim 10^{-13}$ M$_\oplus$ during its 1997 passage near the Sun (mostly at its perihelion). This means that 20 such comets could produce together a CO mass comparable to
that we predicted for the KB. However, the CO comet production is local, anisotropic and concentrated near the Sun. Furthermore, after each comet passage, CO is quickly blown out by the strong Solar wind and escapes at a speed of several au per month. So this CO source cannot in the long run accumulate and compete with the CO production from the Kuiper belt.

Centaurs are transient bodies with a dynamical lifetime of $\sim 10^6-10^7$ yr located between Jupiter and Neptune \citep{2003AJ....126.3122T}. They are expected to originate in the KB \citep{2003AJ....126.3122T}. In spite of their large $\sim$100 km size, the observable Centaurs appear CO-depleted compared to Oort-cloud comets by a factor 10 to 50 \citep{2017AJ....153..230W}. However, there are more than $10^5$ Centaurs larger than about 4 km \citep{2019AJ....158..132N} that may still release CO and contribute to a global and diffuse CO gas disk mainly between 5-30 au in addition to the CO released in the KB. An order of magnitude of the CO gas quantity that could be released by Centaurs in steady state can be obtained as follows. Using the size distribution given by \citet{2019AJ....158..132N}, and integrating over all bodies between 4 and 50 km diameter (OSSOS observation range and in agreement with our model that only bodies larger than 4 km may still have CO in sub-surface), we find that Centaurs have a total surface area of $3.4 \times 10^7$ km$^2$, which is $3\times10^3$ larger than the area of comet Hale-Bopp. 

Assuming that the release rate of Centaurs per unit area is roughly 10 times lower than Hale-Bopp \citep[and it can be up to 50 times,][]{2017AJ....153..230W}, using that the CO release rate of Hale-Bopp at 10 au (the mean distance of Centaur perihelia is close to 13 au but closer in for active centaurs) is $10^{28}$ mole s$^{-1}$ or $2\times 10^{-9}$ M$_\oplus$/Myr \citep{2002EM&P...90....5B}, and assuming that roughly 10\% of Centaurs are active \citep{2012AJ....144...97G}, we find a mean CO outgassing rate of $6\times 10^{-8}$ M$_\oplus$/Myr for Centaurs (or $\sim$ $10^{-8}$ M$_\oplus$/Myr if  the release rate of Centaurs is assumed to be 50 times lower than Hale-Bopp at  the same distance, rather than 10 times). This order of magnitude shows that Centaurs could potentially equally contribute to the CO mass loss rate as do planetesimals in the KB as derived in this paper.

However, the CO gas released by Centaurs is also blown out by the SW much faster than in the KB (e.g., more than 20 times because of the increased proton density at $\sim$10 au), hence decreasing the total mass or column density of CO as compared to that in the KB. A complete modelling of the CO gas released by Centaurs and its evolution is complex and beyond the scope of this paper. However, we also note that the two sources of CO gas could be differentiated from their different spatial distributions by measuring the column at different distances from the Sun (e.g., when an in-situ mission similar to New Horizons with increased sensitivity moves outwards to the KB, see next section).

\section{Detecting the gas belt}\label{obs}

Conceivably, there are different ways of detecting this gas around the Kuiper belt. It may not be exhaustive but the obvious possibilities would be: 1) detection of CO emission at mm-wavelength; 2) detection in the UV (e.g. a resonant carbon line) in absorption against a background star located in the ecliptic; 3) in-situ detections with future missions similar to New Horizons.

\begin{figure*}
   \centering
   \includegraphics[width=14cm]{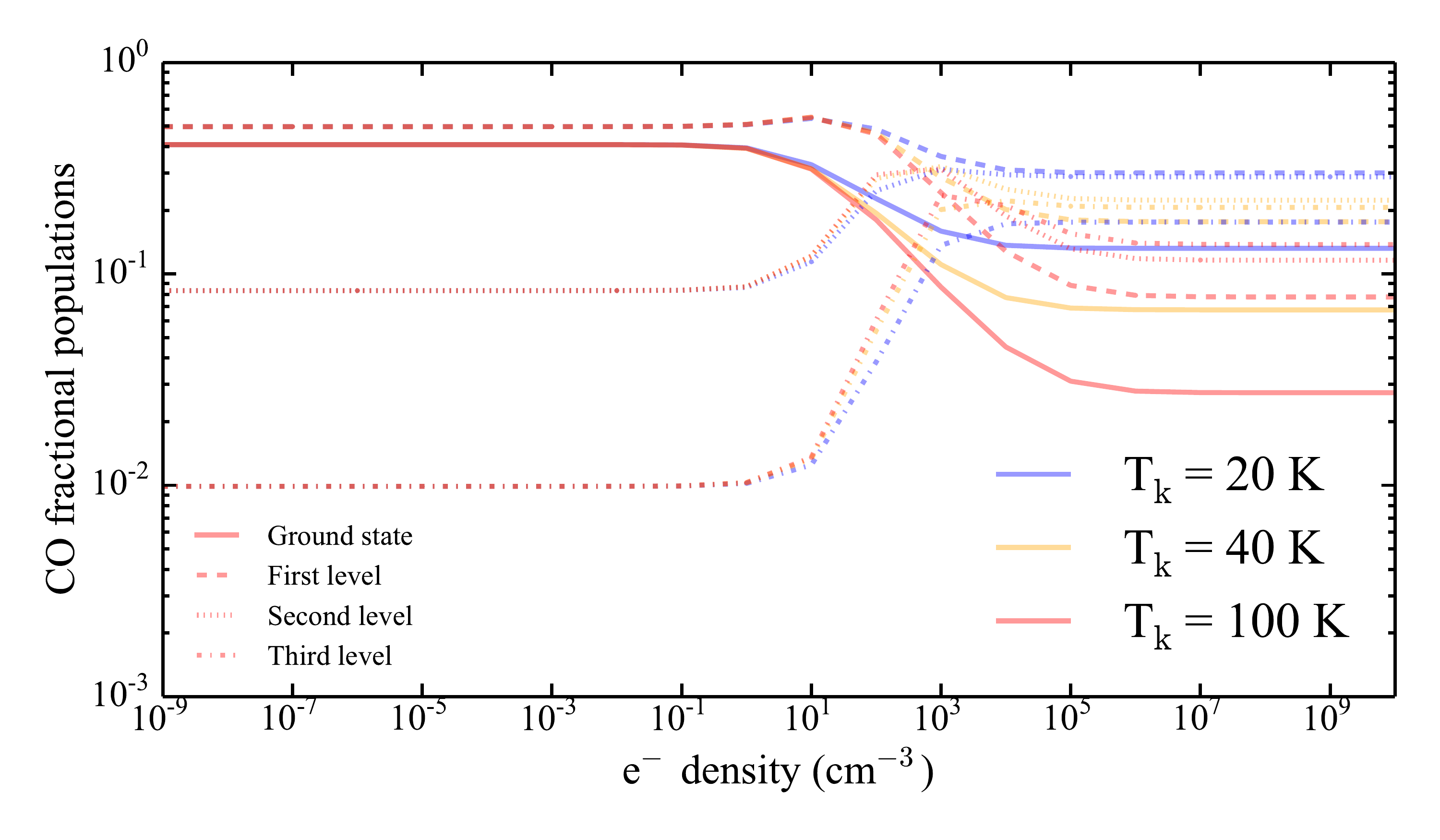}
   \caption{\label{figpop} Population rotational J-levels of CO for a range of temperatures and electron densities. The left side shows the radiation dominated regime (which is the most likely regime for the KB) while the right side has enough colliders to reach LTE.}
\end{figure*}

First, we need to evaluate the population levels for the different lines. Due to the low quantity of electrons we find, the collider density is probably not enough to reach local thermal equilibrium (LTE) and the population levels would be set by the radiation impinging onto the different species. We use a non-LTE code developed for gas in debris disks \citep[including fluorescent excitation,][]{2015MNRAS.447.3936M,2018ApJ...853..147M}. For the radiation field, we include the CMB and the light from the Sun using a state-of-the-art Solar spectrum \citep{2018SoEn..169..434G}. As can be seen in Fig.~\ref{figpop}, we explore the population levels for a range of excitations and temperatures given the current uncertainties on these. For electron densities below $\sim$ 10 cm$^{-3}$ (which is the most likely given our model), the population levels are in the radiation regime and they converge to a given value. What is clear from this plot is that (no matter what the gas temperature is) roughly half of the CO molecules are in the first level and 40\% in the ground state. We will use these values to make flux predictions for different lines below.

Now, let us go through the different possibilities. 

\subsection{Detectability of CO rotational lines in the mm}
\hfill

The Planck mission was used to make Galactic maps of CO using its great sensitivity \citep{2014A&A...571A..13P}. Let us check whether any CO gas in the KB could affect these detections and/or whether CO could be seen on Planck maps looking towards the ecliptic. The Planck sensitivity in band 3 (CO J=1-0 transition at 115 GHz) is roughly 1 K km/s \citep{2014A&A...571A..13P}.

Using the population levels we derived, we compute the column density of CO that is needed to get a 3$\sigma$ detection with Planck \citep[with eq.~9 of][]{1999ApJ...517..209G}. We find that $2 \times 10^{15}$ cm$^{-2}$ of CO is needed to get a detection, which is orders of magnitude higher than the CO column predictions ($\sim 10^{8}$ cm$^{-2}$) from our model. The ALMA non-interferometric total power array mode is more sensitive than Planck given its larger collecting area with 4 $\times$ 12 m antennas. Pushing it to its limit, we find that observing the KB for 1000 hours (as of now, 3000 hr per year are devoted to this mode), we would go down to a sensitivity of 0.45 mK km/s, which would lead to a detection for a CO column density of $10^{12}$ cm$^{-2}$, which is still much higher than our CO column predictions. However, using the auto-correlation mode of ALMA with 50 antennas \citep{2020NatAs...4..861C}, the sensitivity could be 50/4=12.5 times smaller and the detection threshold would then be $\sim 10^{11}$ cm$^{-2}$. Future arrays connecting more numerous and larger antennas in their non-interferometric mode or using autocorrelation (total power) spectra of interferometric data as recently done for observing comet tails with ALMA \citep{2020NatAs...4..861C} would allow to go much deeper. However, we note that we would need to be able to go on and off on the target, which may be difficult given the extent of the KB.

\subsection{Detectability of carbon atoms in the UV in absorption against a background star}
\hfill

We now quantify the absorption signal that would be obtained observing a bright background star that would happen to lie nearby in the ecliptic plane (i.e., the line-of-sight would go through the KB). We take a bright Sirius-like star at a few pc for the example. To estimate the detectability of gas in the KB, we target a strong C I resonant line at 1656.9284 Angstroms as other lines will be of similar or lower strength. We take the Einstein coefficients from the NIST database and compute the optical depth of the line \citep[using eq.~3 of][]{2017MNRAS.464.1415M}. We assume a FWHM of the line of about 10 km/s and using the same non-LTE code as described for CO, we find that the ground state level for carbon is populated at the 99.9\% level. We then find an optical depth $\tau_\nu$ of order $10^{-5}$ for this line.

For an optically thin gas, the flux density of the line is $I_\nu=I_{\nu,{\rm bkg}} \exp(-\tau_\nu)$, where $I_{\nu,{\rm bkg}}$ is the background star flux density. The signal-to-noise ratio of the star (SNR) is $I_{\nu,{\rm bkg}}/\sigma$, where $\sigma$ is the noise level. For a line detection, we need 
$I_{\nu,{\rm bkg}}-I_\nu > 3 \sigma$ at the line centre, which implies an SNR $> 3/\tau_\nu=3 \times 10^{5}$.



To check whether HST/STIS could detect such a faint signal, we use their Exposure Time Calculator. For a Sirius-like star, we find that the star is too bright to be observed directly with STIS. After using an ND filter, we find that we can reach a S/N of order 100 (for 2 hours exposure). However, to get a detection here, we would need a S/N of order $10^5$, which is too much for HST and could only be tackled by future instruments.



\subsection{Detectability of gas with an in-situ mission}
\hfill 

Our Solar System has the advantage over exoplanetary systems in that we can send probes to study its complexity. One such probe is the recent New Horizons mission, which was dedicated to study Pluto (and its satellites) as well as a Kuiper belt object (named Arrokoth).

To detect CO with these in-situ missions, the way to go is to look for absorption of species against the Sun or to look for emission of resonance lines \citep{2016Sci...351.8866G}. For absorption of CO, its ground state has a complex absorption cross-section at wavelengths $<1000$ Angstroms \citep{1980JCP....77..623M} at around $10^{-17}$ cm$^{2}$. With an equivalent of the Alice (UV imaging spectrometer) instrument that is onboard New Horizons, a drop in brightness of $\sim$ 1\% could be detected for a long exposure \citep{2021Icar..35613973G}. Therefore, CO column densities between the instrument and the Sun of the order of $10^{15}$ cm$^{-2}$ may be detected with this technique. This is clearly not enough to detect the CO level predicted by our model.

For some species (mostly atoms), resonance cross-sections can be several orders of magnitude larger than for absorption, which can allow detections of much lower levels of gas in the KB. We now quantify the emission from resonance line scattering for the OI triplet at 1304 Angstrom. To get an order of magnitude of the oxygen upper limit, we consider that there is no background emission and a detection with a signal-to-noise ratio of 3 would require about $C=10$ counts in the wavelength bin of the emission. For a brightness $I=g N$ (in photons/cm$^2$/s/($4\pi$ sr), with $g$ the number of photons scattered per unit time and per atom and $N$ the column density of neutral oxygen), the number of counts in a time $t$ is equal to \citep{1991SSRv...58....1M} $$C = g  N  (\Omega / (4 \pi))  A_{\rm eff}  t,$$ where $\Omega$ is the smallest of the solid angles of the target or detector pixel, and $A_{\rm eff}$ the effective area of the instrument (the aperture area times all the yields and reflectivites) at a given wavelength \citep{2008SSRv..140..155S}. The pixel size of, e.g., the Alice instrument onboard New Horizons is much smaller than the KB gas disk solid angle so that we use the Alice pixel size of $0.1 \times 0.3$ degrees$^2$ to get $\Omega=9.1 \times 10^{-6}$ sr. At the wavelength of the OI 1304 triplet, the effective area is $A_{\rm eff}=0.17$ cm$^2$ \citep{2008SSRv..140..155S}. The $g$ factor \citep[see Table IV of][]{1991SSRv...58....1M} is rescaled at 45 au to get $g=9.4 \times 10^{-10}$ s$^{-1}$. Integrating the number of counts for 1000 h, we get the upper limit $N=2.4 \times 10^{10}$ cm$^{-2}$. It is indeed much more promising than for absorption but given our column density predictions for neutral oxygen ($2 \times 10^{7}$ cm$^{-2}$), this is not doable for now with e.g. Alice. Only future in-situ instruments with larger effective apertures and larger pixel sizes could reduce this time: a super-Alice is needed.

For instance, if we use a 7 deg $\times$ 0.3 deg MCP (micro channel plate) for the detector instead of the current 7 deg $\times$ 0.1 deg, we gain a factor 3, with a resolution that is still sufficient to distinguish lines of interest (e.g., OI at 130.4 nm). Summing over all pixels, instead of just on the $0.1 \times 0.3$ degree$^2$ pixel mentioned above (which is possible because the emission region is larger than the MCP size), the gain factor becomes 70 as $\Omega$ becomes $6.4 \times 10^{-4}$. The effective area $A_{\rm eff}$ can also be improved by a factor $\sim3$ using MgF2-coated optics \citep[as on the Juno mission, which possess an Alice-like spectrograph,][]{2011SPIE.8146E..04D,2013SPIE.8859E..0TG}. All together, and without increasing the aperture size, we gain a factor 200. Using a larger primary would also allow to collect photons faster and go down in sensitivity. For instance, going to a 6 times larger primary (24x24cm) would gain another factor 36 \citep[current airglow aperture is $4 \times 4$ cm$^2$,][]{2008SSRv..140..155S}, i.e., in total a $\sim$7000 gain factor, which is enough (with some slack to account for model uncertainties) to detect the oxygen predicted by our model in the KB with a reasonable exposure time. This type of super-Alice instrument could be planned with current technology and may fly in the future. We also note that if planetesimals in the KB also contain O$_2$ in similar quantity to CO as may be the case in the comet 67 P, our overall prediction for the OI line would also increase by a factor of a few.

\end{appendix}

\label{lastpage}

\end{document}